\documentclass[aps,preprint,amsmath,amssymb,nofootinbib,11pt]{revtex4}
\usepackage{graphicx}
\usepackage{epstopdf}
\usepackage[dvipsnames]{xcolor}
\pdfoutput=1 
\usepackage[T1]{fontenc}

\begin{document}

\title{Sensitivity of anomalous quartic gauge couplings via $Z\gamma\gamma$ production at future hadron-hadron colliders}
\author{A. Senol}
\email[]{senol_a@ibu.edu.tr} 
\author{O. Karadeniz}
\email[]{ozgunkdz@gmail.com} 
\author{K. Y.  Oyulmaz}
\email[]{kaan.oyulmaz@gmail.com} 
\author{C. Helveci}
\email[]{helvaciceren@gmail.com}
\author{H. Denizli}
\email[]{denizli_h@ibu.edu.tr}
\affiliation{Department of Physics, Bolu Abant Izzet Baysal University, 14280, Bolu, Turkey}
\begin{abstract}
Triple gauge boson production provides a promising opportunity to probe the anomalous quartic gauge couplings in understanding the details of electroweak symmetry breaking at future hadron-hadron collider facilities with increasing center of mass energy and luminosity. In this paper, we investigate the sensitivities of dimension-8 anomalous couplings related to the $ZZ\gamma\gamma$ and $Z\gamma\gamma\gamma$ quartic vertices, defined in the effective field theory framework, via $pp\to Z\gamma\gamma$ signal process with Z-boson decaying to charged leptons at the high luminosity phase of LHC (HL-LHC) and future facilities, namely the High Energy LHC (HE-LHC) and Future Circular hadron-hadron collider (FCC-hh). We analyzed the signal and relevant backgrounds via a cut based method with Monte Carlo event sampling where the detector responses of three hadron collider facilities, the center-of-mass energies of 14, 27 and 100 TeV with an integrated luminosities of 3, 15 and 30 ab$^{-1}$ are considered for the HL-LHC, HE-LHC and FCC-hh, respectively. The reconstructed 4-body invariant mass of $l^+l^-\gamma\gamma$ system is used to constrain the anomalous quartic gauge coupling parameters under the hypothesis of absence of anomalies in triple gauge couplings. Our results indicate that the sensitivity on anomalous quartic couplings $f_{T8}/\Lambda^{4}$ and $f_{T9}/\Lambda^{4}$ ($f_{T0}/\Lambda^{4}$, $f_{T1}/\Lambda^{4}$ and $f_{T2}/\Lambda^{4}$) at 95 \% C.L. for FCC-hh with $L_{int}$ = 30 ab$^{-1}$ without systematic errors are two (one) order better than the current experimental limits. Considering a realistic systematic uncertainty such as 10\% from possible experimental sources, the sensitivity of all anomalous quartic couplings gets worsen by about 1.2\%, 1.7\% and 1.5\%  compared to those without systematic uncertainty for HL-LHC, HE-LHC and FCC-hh, respectively.
\end{abstract}

\maketitle

\section{Introduction}\label{secI}
The Standard Model (SM) puzzle was completed with the simultaneous discovery of the scalar Higgs boson, predicted theoretically in the SM, at the CERN Large Hadron Collider (LHC) by both ATLAS and CMS collaborations \cite{Aad:2012tfa,Chatrchyan:2012xdj}. With the discovery of this particle, the mechanism of electroweak symmetry breaking (EWSB) has become more important and still continues to be investigated. The self-interaction of the triple and quartic vector boson couplings is defined by the non-Abelian structure of the ElectroWeak (EW) sector within the framework of the SM. Any deviation in the couplings predicted by the EW sector of the SM is not observed yet with the precision measurements. While the experimental results are consistent with the couplings of $W^{\pm}$ to $Z$ boson, there is no experimental evidence of Z bosons coupling to photons. Therefore, studying of triple and quartic couplings can either confirm the SM and the spontaneous symmetry breaking mechanism or provide clues for the new physics Beyond Standard Model (BSM).  Anomalous triple and quartic gauge boson couplings are parametrized by higher-dimensional operators in the Effective Field Theory (EFT) that can be explained in a model independent way of contribution of the new physics in the BSM. The anomalous triple gauge couplings are modified by integrating out heavy fields whereas the anomalous Quartic Gauge Couplings (aQGC) can be related to low energy limits of heavy state exchange. In this scenario, the $SU(2)_L\bigotimes U(1)_Y$ is realized linearly and the lowest order Quartic Gauge Couplings are given by the dimension-eight operators \cite{Belanger:1992qh, Eboli:2003nq}. These operators are so-called genuine QGC operators which generate the QGC without having TGC associated with them.

Many experimental and phenomenological studies have been carried out and revealed constraints about aQGC. Both vector-boson scattering processes (i.e. $ZZjj$ and $Z\gamma jj$ process ) and triboson (i.e, $Z\gamma\gamma$ production) production are directly sensitive to to the quartic $ZZ\gamma\gamma$ and $Z\gamma\gamma\gamma$ vertices. The  $ZZ\gamma\gamma$ and $Z\gamma\gamma\gamma$ quartic vertices in the framework of SM are forbidden at the tree level and can only be represented by loops in higher order diagrams.
Thus, the anomalous neutral QGCs parametrizing the physics beyond the SM can improve the cross section for $Z\gamma\gamma$ production and affect the kinematic distributions of the final-state Z boson and photons. Since one can fully reconstruct lepton-photon channel of the $Z\gamma\gamma$ final state and calculate the its invariant mass which is sensitive to BSM triboson contributions, we focus on the $Z\gamma\gamma$ production channel at future hadron-hadron colliders.
First experimentally obtained constraint on the aQGCs via $Z\gamma\gamma$ triboson production has been reported in $e^{+}e^{-}$ collisions at LEP \cite{L3:2002axr, OPAL:2004voc}.  During the LHC Run 1, both CMS and ATLAS Collaborations have studied the $pp\rightarrow Z\gamma\gamma \rightarrow l^{+}l^{-}\gamma\gamma$ production channel using an 8 TeV data sample with an integrated luminosity of 20.3 fb$^{-1}$ \cite{ATLAS:2016qjc, CMS:2017tzy}. The most up-to-date result for Z boson production in association with two photons is presented at a center-of-mass energy of 13 TeV with an integrated luminosity of 137 fb$^{-1}$ in Ref. \cite{CMS:2021jji} by CMS collaboration. Apart from $Z\gamma\gamma$ process, the experimental studies on aQGC via three electroweak vector boson production, vector-boson scattering production of electroweak vector boson pairs have been performed at hadron colliders by the Tevatron  \cite{D0:1999oee, D0:2013rce}, the ATLAS \cite{ATLAS:2014jzl, ATLAS:2015ify, ATLAS:2016snd, ATLAS:2017bon,ATLAS:2016nmw} and CMS collaborations \cite{CMS:2014cdf,CMS:2017zmo,CMS:2017rin,CMS:2016gct,CMS:2017fhs,CMS:2019uys,CMS:2019qfk,CMS:2020fqz,CMS:2020ioi,CMS:2020gfh,CMS:2020ypo,CMS:2021gme}. 
In addition to few phenomenological studies on aQGC via $Z\gamma\gamma$ production channel in the literature \cite{Degrande:2013yda,Gutierrez-Rodriguez:2013eya,ATLAS:2013uod,Kurova:2017int,Kurova:2017mnq}, there are some phenomenological studies on aQGC through the vector-boson-scattering production of electroweak vector boson pairs \cite{Belyaev:1998ih, Eboli:2000ad, Perez:2018kav, Guo:2019agy,Guo:2020lim}, the production of three electroweak vector bosons \cite{Yang:2012vv,Ye:2013psa,Wen:2014mha, Zhu:2020ous, Ari:2021ixv}, photon-induced processes with intact protons \cite{ Kepka:2008yx, Chapon:2009ia,Chapon:2009hh,Gupta:2011be,Fichet:2014uka,Baldenegro:2017aen,Tizchang:2020tqs} at hadron colliders, as well as via different production channel at $e\gamma$, $\gamma\gamma$ \cite{Eboli:1993wg, Eboli:1995gv, Hewett:2001dv, Eboli:2001nb, Senol:2014vta} and lepton colliders \cite{Barger:1988fd, Brunstein:1996fz, Han:1997ht, Eboli:1997qz, Stirling:1999ek, Stirling:1999he, Belanger:1999aw, Gangemi:2000sk, Denner:2001vr, Montagna:2001ej, Koksal:2014yia}.

The new era starting with the novel machine configuration of the LHC and beyond aims to decrease the statistical error by increasing center of mass energy and luminosity in the measurements of the Higgs boson properties as well as finding clues to explain the physics beyond SM. With the configurations of beam parameters and hardware, the upgrade project HL-LHC will achieve an approximately 250 fb$^{-1}$ per year to reach a target integrated luminosity of 3000 fb$^{-1}$ at 7.0 TeV nominal beam energy of the LHC in a total of 12 years \cite{ZurbanoFernandez:2020cco}. The other considered post-LHC hadron collider which will be installed in existing LHC tunnel is HE-LHC that is designed to operate at $\sqrt{s}$= 27~TeV center-of-mass energy with an integrated luminosity of at least a factor of 5 larger than the HL-LHC \cite{Abada:2019ono}. As stated in the Update of the European Strategy for Particle Physics by the European Strategy Group, it is recommended to investigate the technical and financial feasibility of a future hadron collider at CERN with a centre-of-mass energy of at least 100 TeV. The future project currently under consideration by CERN which comes to fore with infrastructure and technology as well as the physics opportunities is the Future Circular Collider (FCC) Study \cite{FCC:2018vvp}.

The goal of our study is to investigate the effects of anomalous quartic gauge couplings on $ZZ\gamma\gamma$ and $Z\gamma\gamma\gamma$ vertices via $pp\rightarrow Z\gamma\gamma$ process where Z boson subsequently decays to $e$ or $\mu$ pairs at HL-LHC, HE-LHC and FCC-hh. The rest of the paper is organized as follows. A brief review of theoretical framework that discusses the operators in EFT Lagrangian is introduced in Section~\ref{secII}. The event generation tools as well as the detail of the analysis to find the optimum cuts for separating signal events from different source of backgrounds is discussed in Section~\ref{secIII}. In section~\ref{secIV}, we give the detail of method to obtain sensitivity bounds on anomalous quartic gauge couplings, and then determine them with an integrated luminosity $L_{int}$ = 3 ab$^{-1}$, 15 ab$^{-1}$, 30 ab$^{-1}$ for HL-LHC, HE-LHC and FCC-hh, respectively. Finally, we summarize our result and compare obtained limits to the current experimental results in Section~\ref{secV}.

\section{Dimension-eight operators for anomalous Quartic $ZZ\gamma\gamma$ and $Z\gamma\gamma\gamma$ couplings}\label{secII}
Although there is no contribution of the quartic gauge-boson couplings of the $ZZ\gamma\gamma$ and $Z\gamma\gamma\gamma$ vertices to the $Z\gamma\gamma$ production in the SM, new physics effects in the cross section of $Z\gamma\gamma$ production can be searched with high-dimensional effective operators which describe the anomalous quartic-gauge boson couplings without  triple gauge-boson couplings. These neutral aQGCs couplings are modeled by either linear or non-linear representations using an EFT \cite{Eboli:2006wa, Degrande:2013rea, Baak:2013fwa}. In the non-linear representation, the electroweak symmetry breaking is due to no fundamental Higgs scalar whereas in the linear representation,  it can be  broken by the conventional SM Higgs mechanism. With the discovery of the Higgs boson at the LHC, it becomes important to study the anomalous quartic gauge couplings based on linear representation. In this representation,  the parity conserved and charge-conjugated effective Lagrangian include the  dimension-eight effective operators by assuming the $SU(2)\times U(1)$ symmetry of the EW gauge field, with a Higgs boson belongs to a $SU(2)_L$ doublet. In this approach, the lowest dimension of operators which leads to quartic interactions but do not include two or three weak gauge boson vertices are expected to be eight.  Therefore, the three classes of dimension-eight effective operators containing; i) covariant derivatives of Higgs doublet only ($\mathcal O_{S,j}$), ii) two field strength tensors and two derivatives of Higgs doublet ($\mathcal O_{M,j}$) and iii) field strength tensors only  ($\mathcal O_{T,j}$) are added SM Lagrangian
\begin{eqnarray}\label{lag}
\mathcal{L}_{eff}=\mathcal{L}_{SM}+\sum_{j=0}^{1}\frac{f_{S,j}}{\Lambda^4}\mathcal{O}_{S,j}+\sum_{j=0}^{7}\frac{f_{M,j}}{\Lambda^4}\mathcal{O}_{M,j}+\sum_{j=0\atop \ j\neq 3}^{9}\frac{f_{T,j}}{\Lambda^4}\mathcal{O}_{T,j}
\end{eqnarray}
where $\Lambda$ is the scale of new physics, and $f_{S,j}$, $f_{M,j}$ and  $f_{T,j}$ represent coefficients of relevant effective operators.  These coefficients are zero in the SM prediction. The expanded form of these operators and a complete list of quartic vertices modified by these operators are given in Appendix~\ref{app}. Among the $f_{M,x}$ and $f_{T,x}$ operators that affect the $ZZ\gamma\gamma$ and $Z\gamma\gamma\gamma$ vertices, $f_{M,x}$ operators in the production of $Z\gamma\gamma$ at future of hadron-hadron colliders with high center of mass energies and luminosities were examined and their limit values were predicted in Ref[29]. Therefore, we focus on the five coefficients $f_{T0}/\Lambda^4$, $f_{T1}/\Lambda^4$, $f_{T2}/\Lambda^4$, $f_{T8}/\Lambda^4$ and $f_{T9}/\Lambda^4$  of the operators containing four field strength tensors for this study. Especially  $f_{T8}/\Lambda^4$, and $f_{T9}/\Lambda^4$ give rise to only neutral anomalous quartic gauge vertices.



\begin{figure}
\includegraphics[scale=1.0]{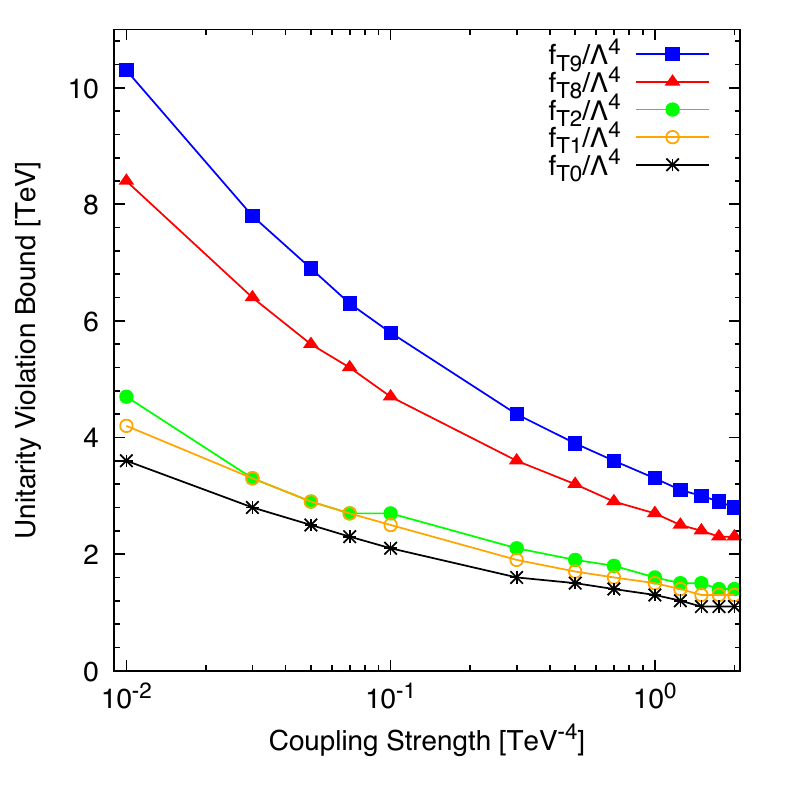} 
\caption{The unitarity violation bound as a function of  $f_{T,j}$ (where $j=0,1,2,8,9$) aQGC via on-shell VV ($V = W / Z / \gamma$) scattering processes. \label{fig1}}
\end{figure}

The effective field theory is only valid under the new physics scale in which unitarity violation does not occur. However, high-dimensional operators with nonzero aQGC can lead to a scattering amplitude that violates unitarity at sufficiently high energy values, called the unitarity bound. The value of the unitarity bound for the dimension-8 operators is determined by using a dipole form factor ensuring unitarity at high energies as: 
\begin{eqnarray}
FF=\frac{1}{(1+\hat s/\Lambda_{FF}^2)^2}
\end{eqnarray}
where $\hat s$ is the maximum center-of-mass energy, $\Lambda_{FF}$ is the energy scale of the form factor. The maximal form factor scale $\Lambda_{FF}$ is calculated with a form factor tool VBFNLO 2.7.1 \cite{Arnold:2008rz} for a given input of anomalous quartic gauge boson couplings parameters. The VBFNLO utility determines form factor using the amplitudes of on-shell VV scattering processes and computes the zeroth partial wave of the amplitude. The real part of the zeroth partial wave must be below 0.5 which is called the unitarity criterion. All channels  the same electrical charge $Q$ in $VV\to VV$ scattering ($V = W / Z / \gamma$)  are combined in addition to individual check on each channel of the $VV$ system.  The calculated Unitarity Violation (UV) bounds using the form factor tool with VBFNLO as a function of higher-dimensional operators considered in our study are given in Fig.\ref{fig1}. The unitarity is safe in the region that is below the line for each coefficients. 
\begin{table}
\caption{The current experimental limits at 95\% CL on the anomalous quartic gauge couplings $f_{T0}/\Lambda^{4}$, $f_{T1}/\Lambda^{4}$, $f_{T2}/\Lambda^{4}$, $f_{T8}/\Lambda^{4}$ and $f_{T9}/\Lambda^{4}$ obtained from analysis of $Z\gamma\gamma$ production as well as different production channels by ATLAS and CMS collaborations. \label{tab1}}
\begin{ruledtabular}
\begin{tabular}{l|c|c|ccc}
Collaboration& ATLAS ($\sqrt s$=8~TeV)& CMS ($\sqrt s$=13 TeV) & \multicolumn{3}{c}{Current Limits by CMS ($\sqrt s$=13 TeV)}\\ 
Luminosity&L$_{int}$=20.3 fb$^{-1}$ &L$_{int}$=137 fb$^{-1}$ & L$_{int}$=35.9 fb$^{-1}$&  \multicolumn{2}{c}{ L$_{int}$=137 fb$^{-1}$}\\
Channel&$Z\gamma\gamma$ \cite{ATLAS:2016qjc}&  $Z\gamma\gamma$ \cite{CMS:2021jji} & $WVjj(V=W,Z)$ \cite{CMS:2019qfk}& $ZZjj$ \cite{CMS:2020fqz} & $Z\gamma jj$ \cite{CMS:2021gme} \\\hline\hline
$f_{T0}/\Lambda^{4}$(TeV$^{-4}$)& [-0.86;1.03]$\times 10^{2}$ &[-5.70;5.46] & [-0.12;0.11] & - & -\\ 
$f_{T1}/\Lambda^{4}$(TeV$^{-4}$)& - & [-5.70;5.46] &[-0.12;0.13] &  & - \\ 
$f_{T2}/\Lambda^{4}$(TeV$^{-4}$)& - & [-11.4;10.9] & [-0.28;0.28] &-  & - \\ 
$f_{T8}/\Lambda^{4}$(TeV$^{-4}$)& - & [-1.06;1.10] &-&  [-0.43;0.43]& - \\ 
$f_{T9}/\Lambda^{4}$(TeV$^{-4}$)& [-0.74;0.74]$\times10^{4}$ & [-1.82;1.82] & - & - &[-0.91;0.91] \\ 
\end{tabular}
\end{ruledtabular}
\end{table}

The limit values with no unitarization restriction ( $\Lambda_{FF}=\infty$) on the dimension-8 aQGC ($f_{T0}/\Lambda^{4}$, $f_{T1}/\Lambda^{4}$, $f_{T2}/\Lambda^{4}$, $f_{T8}/\Lambda^{4}$ and $f_{T9}/\Lambda^{4}$) obtained by ATLAS and CMS collaborations by setting all other anomalous couplings to zero are summarized in Table \ref{tab1}. The limits on $f_{T0}/\Lambda^{4}$, $f_{T1}/\Lambda^{4}$, $f_{T2}/\Lambda^{4}$,$f_{T8}/\Lambda^{4}$ and $f_{T9}/\Lambda^{4}$ related to the quartic $ZZ\gamma\gamma$ and $Z\gamma\gamma\gamma$ vertices have been provided by ATLAS \cite{ATLAS:2016qjc}(CMS \cite{CMS:2021jji}) collaboration at $\sqrt{s}= 8$~TeV (13 TeV) for an integrated luminosity 20.3 fb$^{-1}$ (137 fb$^{-1}$) via two isolated high-energy photons and a $Z$ boson (subsequently, decay to charged leptons) which is the signal process considered in our analysis. 

The current  best limits obtained by CMS collaboration for different production channels on $f_{T0}/\Lambda^{4}$, $f_{T1}/\Lambda^{4}$, $f_{T2}/\Lambda^{4}$, $f_{T8}/\Lambda^{4}$ and $f_{T9}/\Lambda^{4}$ couplings are also presented in last column of the Table \ref{tab1}.
The best limits obtained on $f_{T0}/\Lambda^{4}$, $f_{T1}/\Lambda^{4}$, $f_{T2}/\Lambda^{4}$ from $WW$, $WZ$, and $ZZ$ boson pairs in association with two jets production \cite{CMS:2019qfk} at a center-of-mass energy of 13 TeV with an integrated luminosity 35.9 fb$^{-1}$. They also reported limits on $f_{T8}/\Lambda^{4}$ from production of two jets in association with two Z boson \cite{CMS:2020fqz} and  $f_{T9}/\Lambda^{4}$ from electroweak production of a $Z$ boson, a photon and two forward jets production \cite{CMS:2021gme} at a center-of-mass energy of 13 TeV with an integrated luminosity 137 fb$^{-1}$. 
\section{Event Selection and Details of Analysis}\label{secIII}
\begin{figure}[htb!]
\includegraphics[scale=0.95]{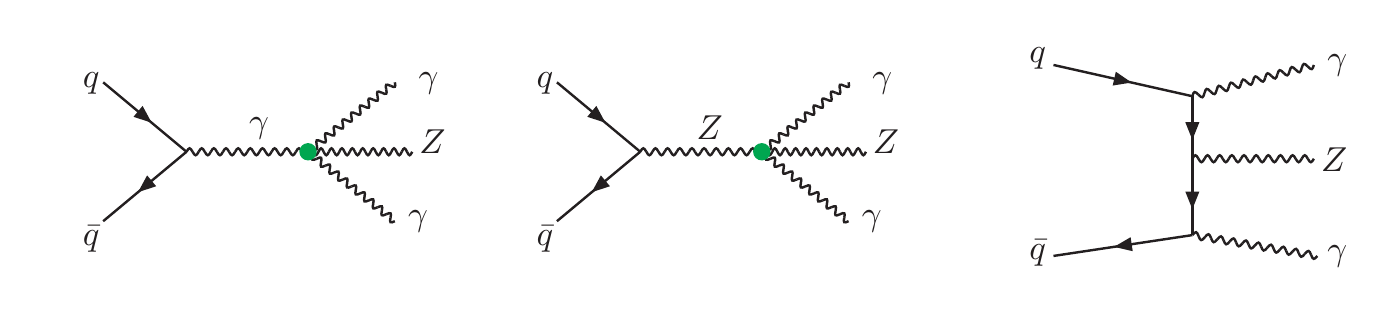}
\caption{Feynman diagrams for the tree-level $Z\gamma\gamma$ production of anomalous quartic gauge-boson coupling vertex (represented with green circle) and the SM contributions. \label{fig2}}
\end{figure}

\begin{figure}[htb!]
\includegraphics[scale=0.69]{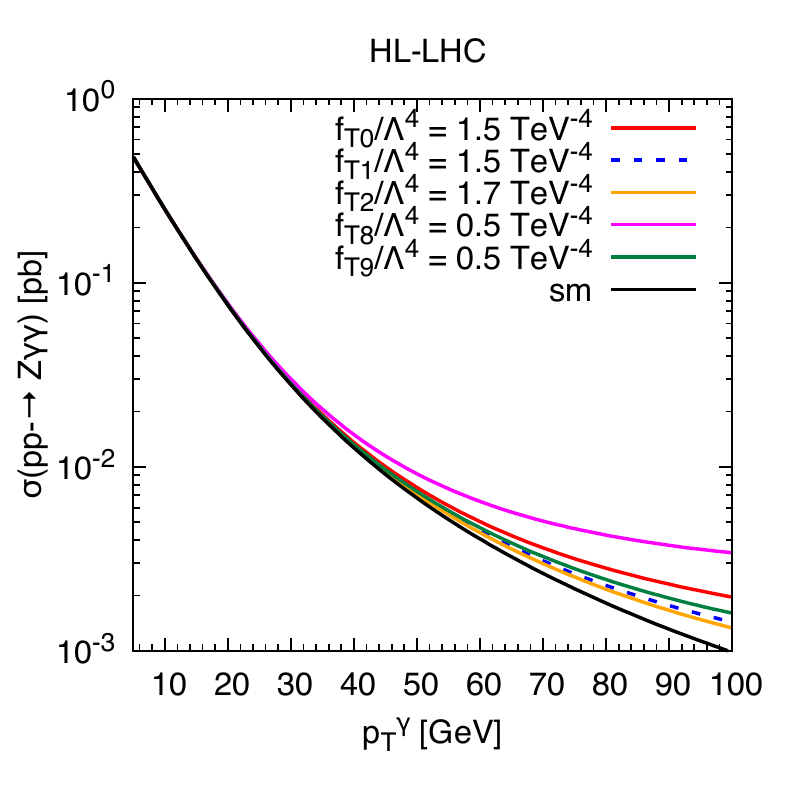}\includegraphics[scale=0.69]{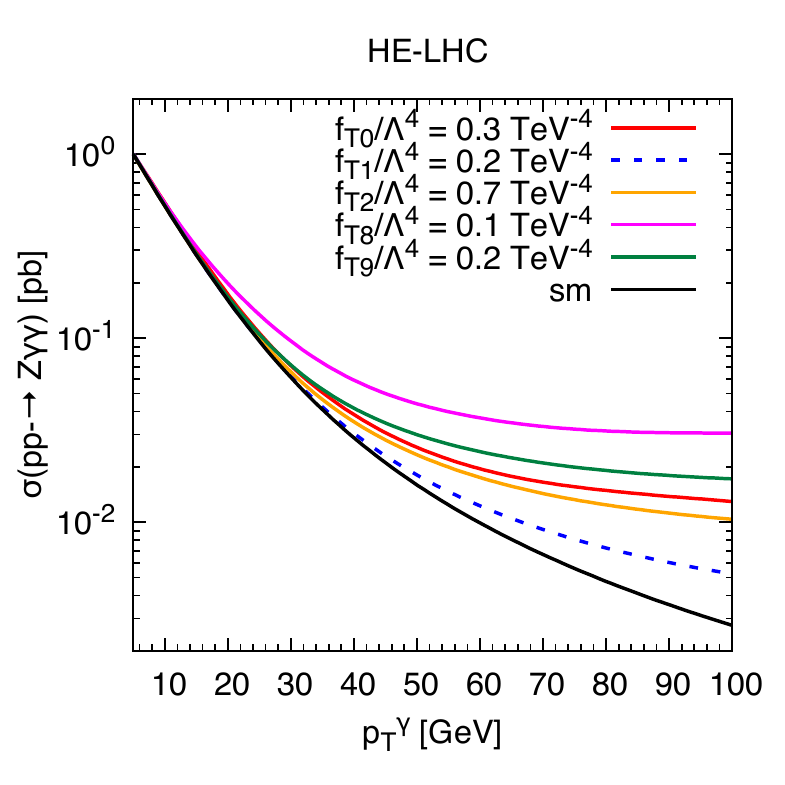}\includegraphics[scale=0.69]{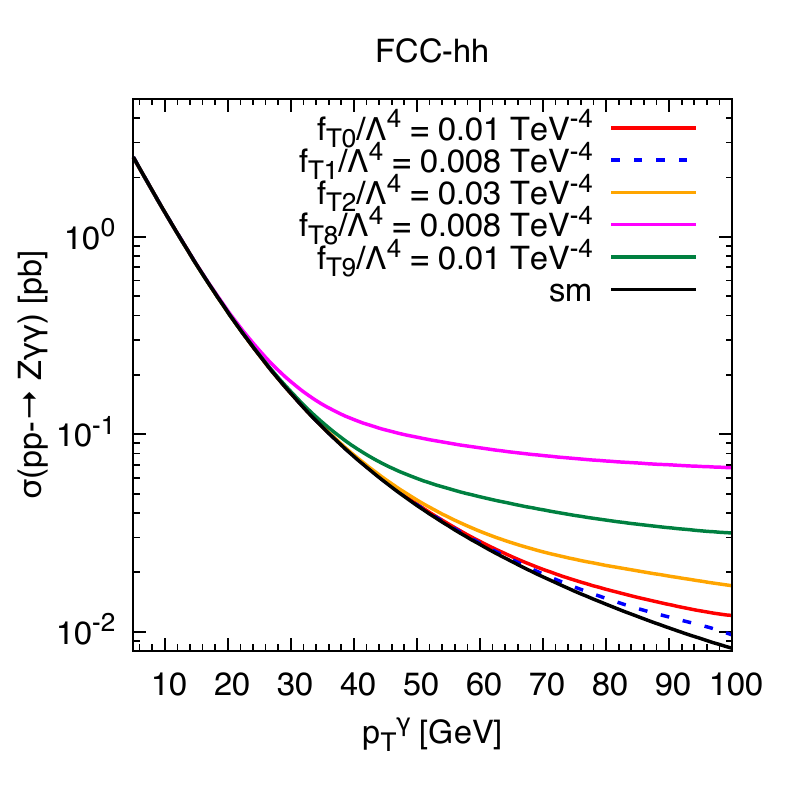}\\  
\includegraphics[scale=0.69]{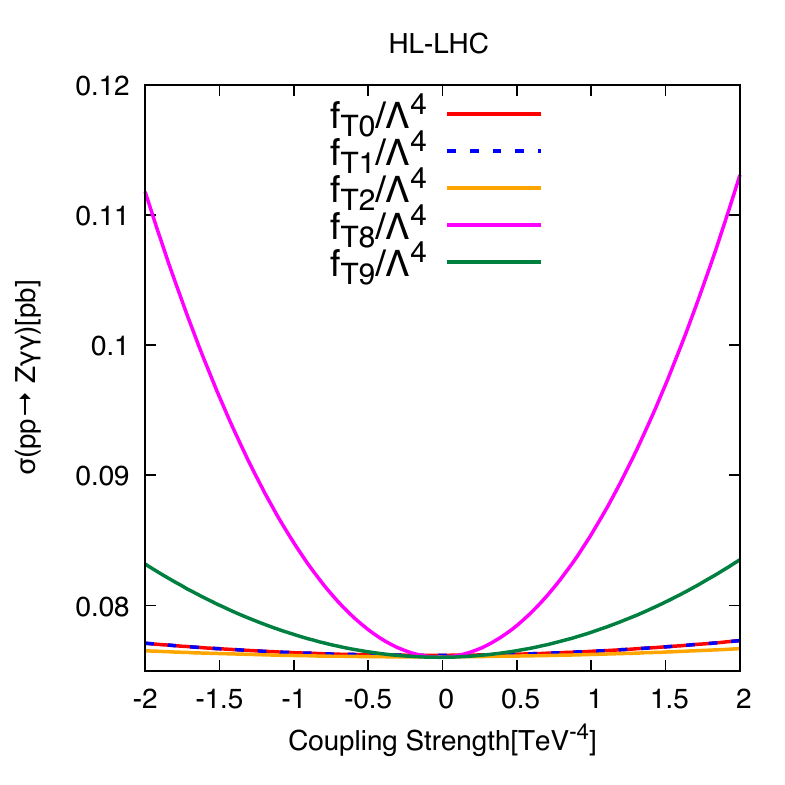}\includegraphics[scale=0.69]{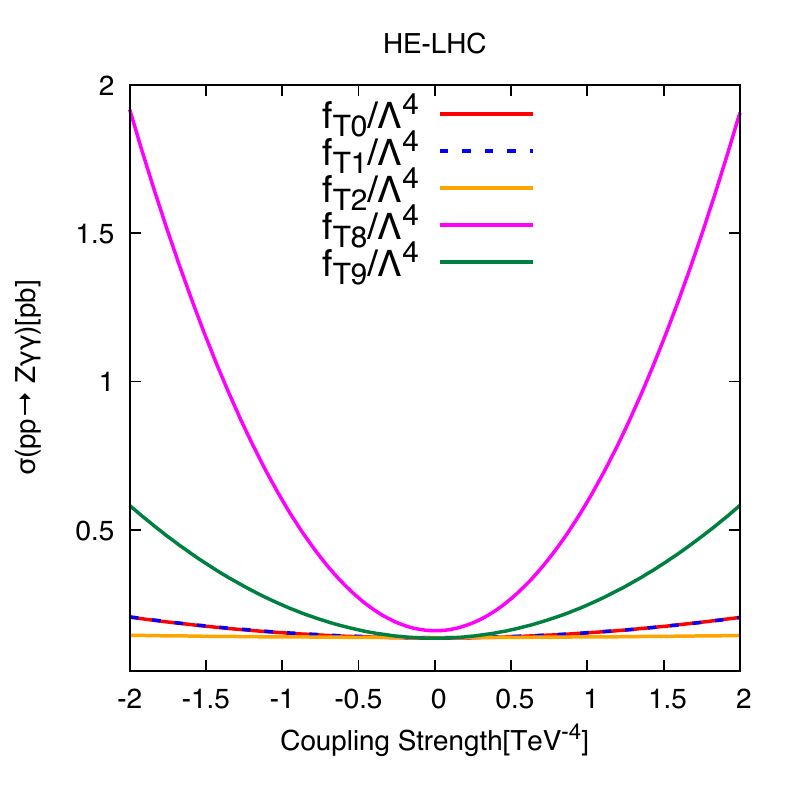}\includegraphics[scale=0.69]{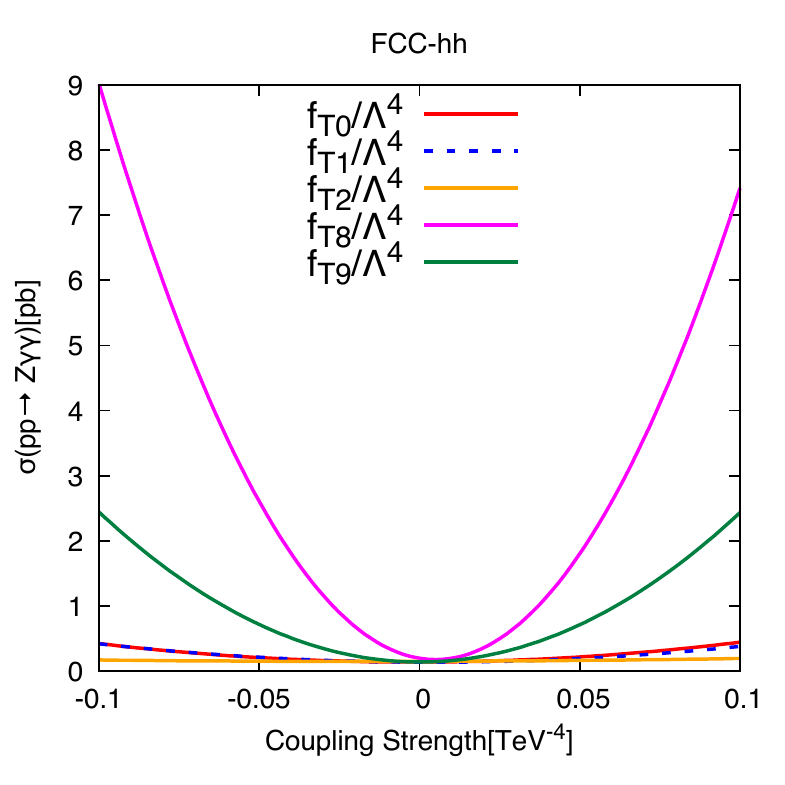}
\caption{In the top row the cross section of the $p p \rightarrow Z\gamma\gamma$ process as function of the lower cut on photon transverse momentum are shown while the bottom row shows production cross section of the $Z\gamma\gamma$ process in terms of the anomalous quartic gauge couplings for HL-LHC, HE-LHC and  FCC-hh (left to right, respectively.) \label{fig3}}
\end{figure}

\begin{figure}[htb!]
\includegraphics[scale=0.43]{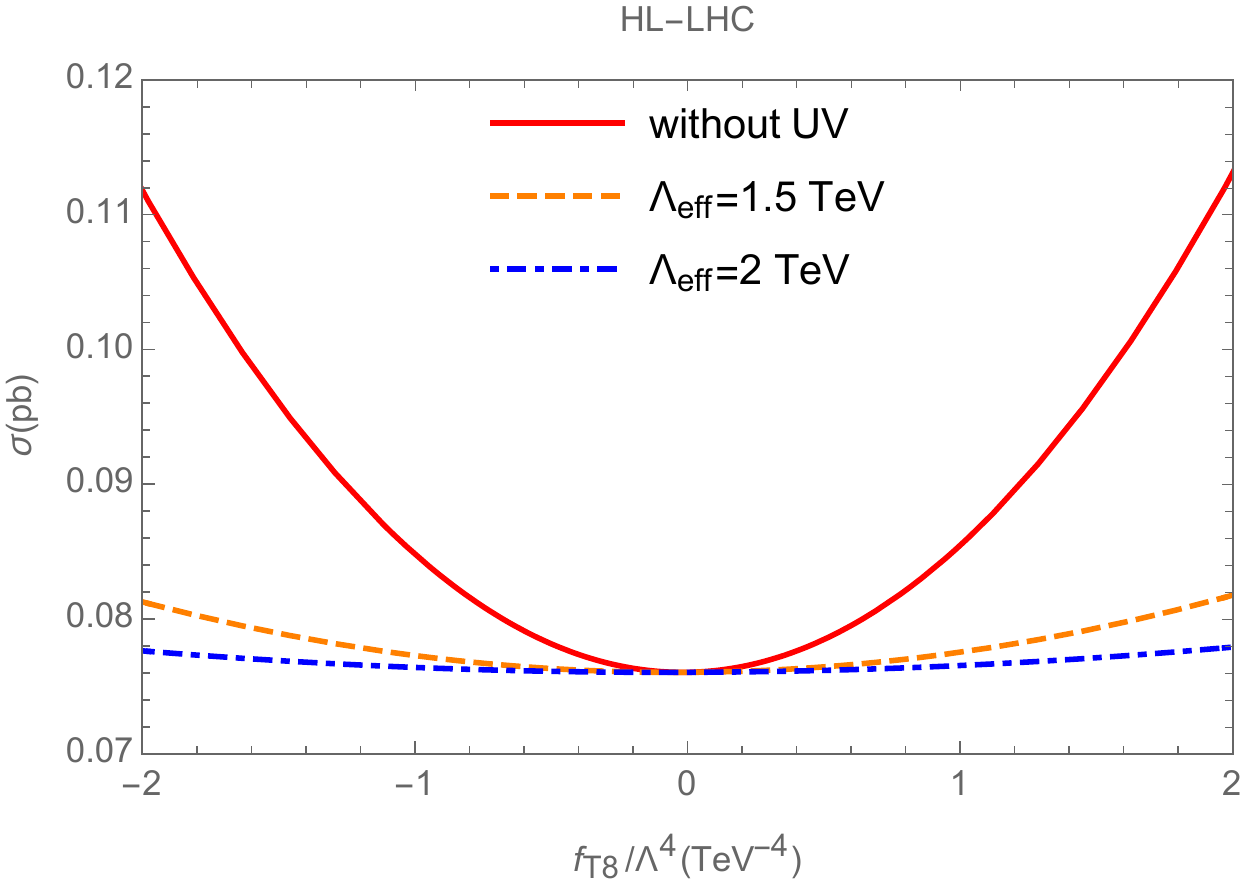}\includegraphics[scale=0.43]{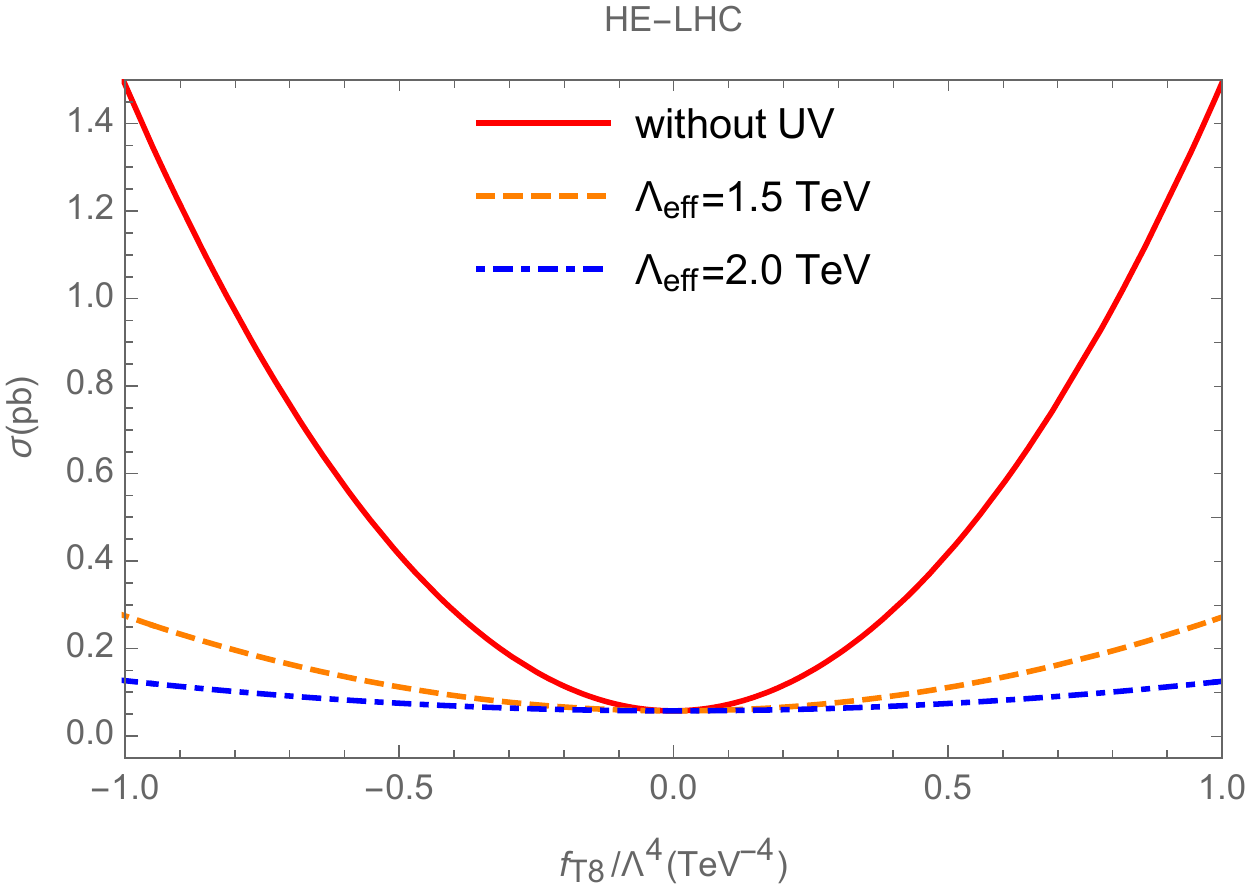}\includegraphics[scale=0.43]{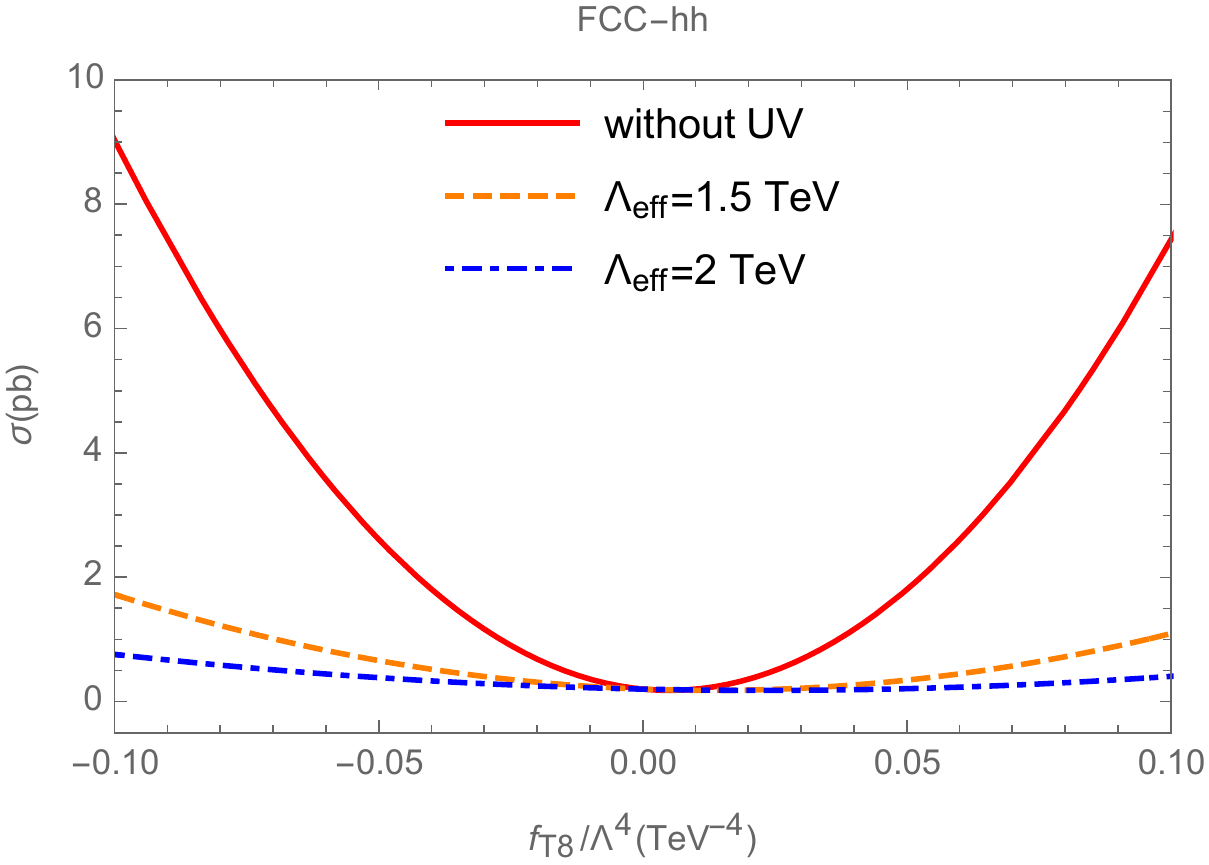}\\  
\caption{The cross section of the $pp \rightarrow Z\gamma\gamma$ process as function of the anomalous quartic gauge coupling $f_{T8}/\Lambda^{4}$ for HL-LHC, HE-LHC and  FCC-hh without UV bound applied and with form factor scale $\Lambda_{FF}$=1.5 and 2 TeV.  \label{fig3_2}}
\end{figure}

The details of the analysis are given for the effects of dimension-8 operators on anomalous quartic gauge boson couplings via $pp\rightarrow Z\gamma\gamma$ signal process for future hadron-hadron circular colliders in this section. The $pp\rightarrow Z\gamma\gamma$  process receives contributions from three Feynman Diagrams at the three-level as shown in Fig.\ref{fig2}. The first two diagrams account for the anomalous quartic $ZZ\gamma\gamma$ and $Z\gamma\gamma\gamma$ couplings while the other represents the SM contribution. The green circles indicate the vertices which are defined by the operators Eqs.(\ref{eq2})-(\ref{eq3}) in the Lagrangian (Eq.\ref{lag}). The cross section and dynamical distributions are evaluated using a parton level Monte-Carlo (MC) simulations which are performed within the {\sc MadGraph5\_aMC@NLO} \cite{Alwall:2014hca}. In addition, the effective Lagrangians of $ZZ\gamma\gamma$ and $Z\gamma\gamma\gamma$ anomalous quartic couplings are implemented into {\sc MadGraph5\_aMC@NLO} based on {\sc FeynRules} \cite{Alloul:2013bka} and Universal {\sc FeynRules} Output (UFO)  framework \cite{Degrande:2011ua}. In our Leading-Order (LO) cross section calculations at the generator level to consider the finite acceptance of the detectors and isolation of the final state observable particles, we applied cuts on pseudo-rapidity of photons as $|\eta^{\gamma}|<2.5$ and on separation of two photon in the pseudorapidity-azimuthal angle plane as $\Delta R(\gamma,\gamma)>0.4$. Additionally, we introduced a generator level cut on the transverse momentum of the photons not only to guarantee that our results are free of infrared divergences but also to improve event generating efficiency. In order to determine transverse momentum cut on final state photons at generator level, the cross section of $Z\gamma\gamma$ process as a function of the lower cut on photon transverse momentum are plotted  for HL-LHC ($\sqrt{s}$=14~TeV), HE-LHC($\sqrt{s}$=27~TeV) and FCC-hh ($\sqrt{s}$=100~TeV) colliders in the first row of Fig \ref{fig3}, respectively. 
The cross sections with one non-zero anomalous quartic gauge couplings at a time as well as SM are selected in these figures for the illustrative purpose. As seen from these figures, $p_{T}^\gamma > 15$ GeV for HL-LHC and HE-LHC, $p_{T}^\gamma > 25~GeV$ for FCC-hh are chosen for further analysis. The second row of Fig.\ref{fig3} shows the production cross section of the $Z\gamma\gamma$ process in terms of the anomalous quartic gauge couplings for three different hadron-hadron collider options with corresponding generator level cuts. Each line in the second row of Fig. \ref{fig3}  corresponds to variation of cross section with respect to one of the coupling while others are kept at zero. The sensitivity of $f_{T8}/\Lambda^{4}$ and $f_{T9}/\Lambda^{4}$ anomalous quartic couplings for each collider options to cross sections is more significant than the other couplings as seen in the second row of Fig. \ref{fig3}. Therefore, we expect to obtain better limits on the $f_{T8}/\Lambda^{4}$ and $f_{T9}/\Lambda^{4}$ couplings with  $Z\gamma\gamma$ production. The general expression for amplitude in the EFT regime for the process considered can be written as
\begin{eqnarray}
\resizebox{0.50\textwidth}{!}{
$\left|M_{SM}+M_{dim8}\right|^2 = \underbrace{\left|M_{SM}\right|^2}_{\mathcal{O}\left(\Lambda^0\right)}+\underbrace{2\Re\left(M_{SM}M_{dim8}^*\right)}_{\mathcal{O}\left(\Lambda^{-4}\right)}+\underbrace{\left|M_{dim8}\right|^2}_{\mathcal{O}\left(\Lambda^{-8}\right)}$
}
\end{eqnarray}
where $\left|M_{SM}\right|^2$, $\Re\left(M_{SM}M_{dim8}^*\right)$ and $\left|M_{dim8}\right|^2$ are the SM, interference of the SM amplitude with higher dimensional operators, and the square of the new physics contributions, respectively. In order to show the effectiveness of the form factor, the cross sections at LO without and with $\Lambda_{FF}$=1.5 and 2 TeV  is presented for all three collider options in Fig.\ref{fig3_2}.  It can be clearly seen in Fig. \ref{fig3_2} that the square contributions of the new physics amplitudes suppress the interference contributions of the SM amplitude with high-dimensional operators in the case where the UV limit is not applied. However, if the new physics energy scale is heavy (i.e. $\Lambda_{FF}$=1.5 and 2 TeV or higher), the largest new physics contribution to $pp \rightarrow Z\gamma\gamma$ process is expected from the interference between the SM and the dimension-eight operators as seen from Fig. \ref{fig3_2}.

For further analysis including response of the detector effects, we generate 600k events for all backgrounds and signal processes where we scan each $f_{T0}/\Lambda^{4}$, $f_{T1}/\Lambda^{4}$, $f_{T2}/\Lambda^{4}$, $f_{T8}/\Lambda^{4}$ and $f_{T9}/\Lambda^{4}$ couplings with nine different values ranging from negative to positive. The generated signal and (relevant) background events with applied generator level cuts are first passed through {\sc Pythia 8.20} \cite{Sjostrand:2014zea} for hadronization, parton showering and decay of unstable particles at parton level. Then hadronized events are interfaced with {\sc Delphes 3.4.2} \cite{deFavereau:2013fsa} software to model response of the corresponding detector for each collider. 
We choose the upgraded cards namely \verb|delphes_card_HLLHC_aqgc.tcl|  and \verb|FCC-hh.tcl| where detector response in the form of resolution functions and efficiencies are parametrised for  HL-LHC, HE-LHC and FCC-hh, respectively. Jets are reconstructed by using clustered energy deposits with {\sc FastJet 3.3.2} \cite{Cacciari:2011ma} using anti-kt algorithm \cite{Cacciari:2008gp} where a cone radius is set as $\Delta R$ = 0.4 (0.2)  and $p_T^j$>15 (25) GeV for  HL-LHC and HE-LHC (FCC-hh) colliders.


Our main focus is to see the effects of anomalous quartic gauge boson couplings via $pp \rightarrow Z\gamma\gamma$ signal process where $Z$ boson subsequently decays to $e$ or $\mu$ pairs. Therefore, events with two isolated photons and one pair of the same flavor and oppositely charged leptons (electrons or muons) are selected for further analysis (Cut-0). Electron and muon channels are combined to increase sensitivity even more. The signal includes nonzero effective couplings and SM contribution as well as its interference. 
"sm" stands for SM background process of the same final state with the signal process in our analysis. The main background processes to the selected $l^{+}l^{-}\gamma\gamma$ sample of events may originate from $Z\gamma j$ and $Zjj$ production with hadronic jet misidentified as a photon. Such misidentifications generally arise from jets hadronizing with a neutral meson, which carries away most of the jet energy. The photons that carry a large fraction of the jet energy can exceed the reconstructed photon isolation requirements. The probability of misidentifying a jet as a photon depends on how the jets interact with the detector. Due to the great angular resolution of existing calorimeters in current LHC detectors, the probability of a hard scattering jet being mis-reconstructed as an isolated photon is low. The granularity for the detectors of future hadron-hadron colliders, for which we will use detector parameters, is expected to be 2-4 times better than current LHC detectors. Therefore, we applied the same rate as in the HL-LHC detectors for all colliders options in a fake-rate of jet-photon which can be parametrized in the form 
$\epsilon_{j \rightarrow \gamma}= 0.0007\cdot e^{-p_{T}[GeV]/187}$ \cite{ATLAS:2016ukn,Mangano:2020sao}. The remaining minor background contributions are originated from $WZ$, $ZZ$ and $\tau^+\tau^-\gamma\gamma$ processes. $WZ$ and $ZZ$ backgrounds are considered either the electrons from the decay of the $W$ and $Z$ boson are misidentified as photons or  final state photons are radiated. $\tau^+\tau^-\gamma\gamma$ has also the same final state as the considered signal process.
\begin{table}
\caption{Event selection criteria and applied kinematic cuts for the analysis of the HL-LHC, HE-LHC and FCC-hh collider options \label{tab2}}
\begin{ruledtabular}
\begin{tabular}{lccccccc}
Cuts& HL-LHC & \vline &  HE-LHC &\vline &FCC-hh \\ \hline
Cut-0&&& $N_{l} \geqslant 2$ and same flavour but opposite charge, $N_{\gamma} \geqslant 2$ \\\hline
Cut-1&&& $p_{T}^{l_1,l_2} > 25$ GeV, $| \eta^{l_1,\,l_2}|\leqslant 2.5$  \\\hline
Cut-2&  $p_{T}^{\gamma_1,\gamma_2} >  15$ GeV &\vline& $p_{T}^{\gamma_1,\gamma_2} >  20$ GeV &\vline& $p_{T}^{\gamma_1,\gamma_2} >  20$ GeV \\ & $| \eta^{\gamma_1,\gamma_2}|\leqslant 2.5$ &\vline& $|\eta^{\gamma_1,\gamma_2}|\leqslant 2.5$ &\vline& $|\eta^{\gamma_1,\gamma_2}|\leqslant 2.5$\\\hline
Cut-3&&& $\Delta R ({\gamma_1,\gamma_2}$)> 0.4, $\Delta R({\gamma_1,l_1}$)> 0.4, $\Delta R({l_1,l_2})$< 1.4 \\\hline
Cut-4&&& 81 GeV $ < M_{l^+l^-} < $ 101 GeV \\\hline
Cut-5& $p_{T}^{\gamma_1} > 160$ GeV &\vline& $p_{T}^{\gamma_1} > 250$ GeV &\vline& $p_{T}^{\gamma_1} > 300$ GeV\\
\end{tabular}
\end{ruledtabular}
\end{table}
In our analysis, we will use the traditional cut-based method to obtain the best possible bounds for anomalous quartic gauge couplings. This method is based on determining the optimum kinematic cuts from the kinematic distributions of the final state particles in the selected events and then maximizing the number of signal events with a powerful single kinematic observable that can distinguish between the signal and the background using these cuts. The charged leptons (photons) in the final state are ordered with respect to their transverse momenta; the leading and sub-leading as $l_1$ ($\gamma_1$) and $l_2$ ($\gamma_2$), respectively. All kinematical distributions are normalized to the number of expected events (cross section of each process times the integrated luminosity $L_{int}$ =  3 ab$^{-1}$, 15 ab$^{-1}$, 30 ab$^{-1}$ for HL-LHC, HE-LHC and FCC-hh, respectively). 
\begin{figure}
\includegraphics[scale=0.23]{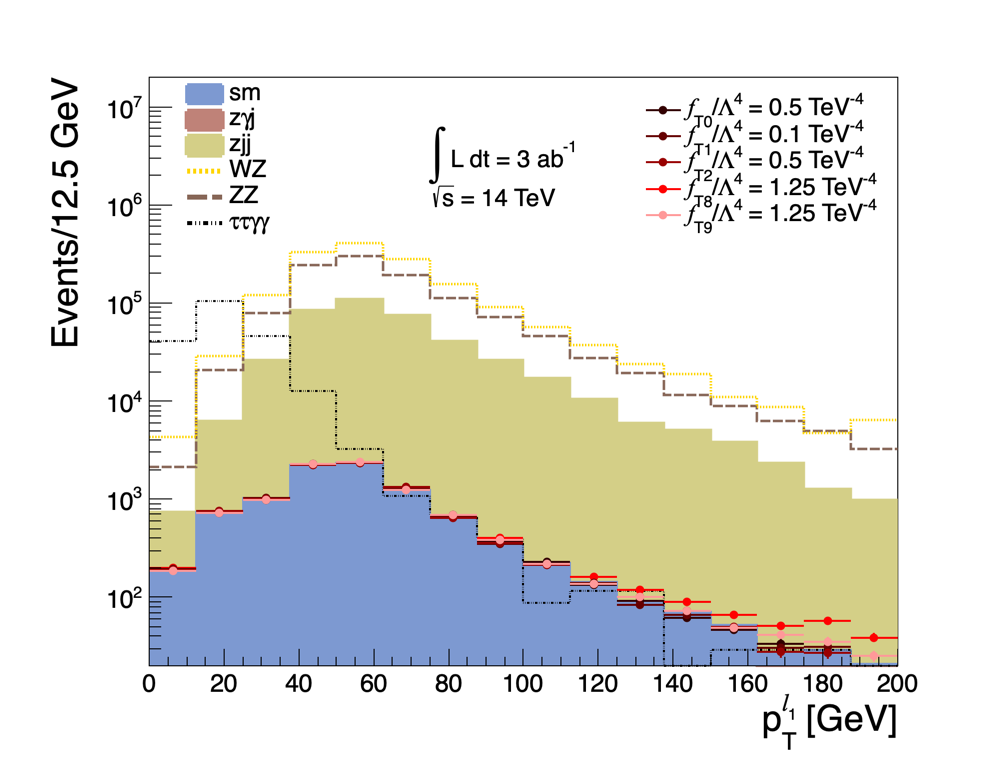}\includegraphics[scale=0.23]{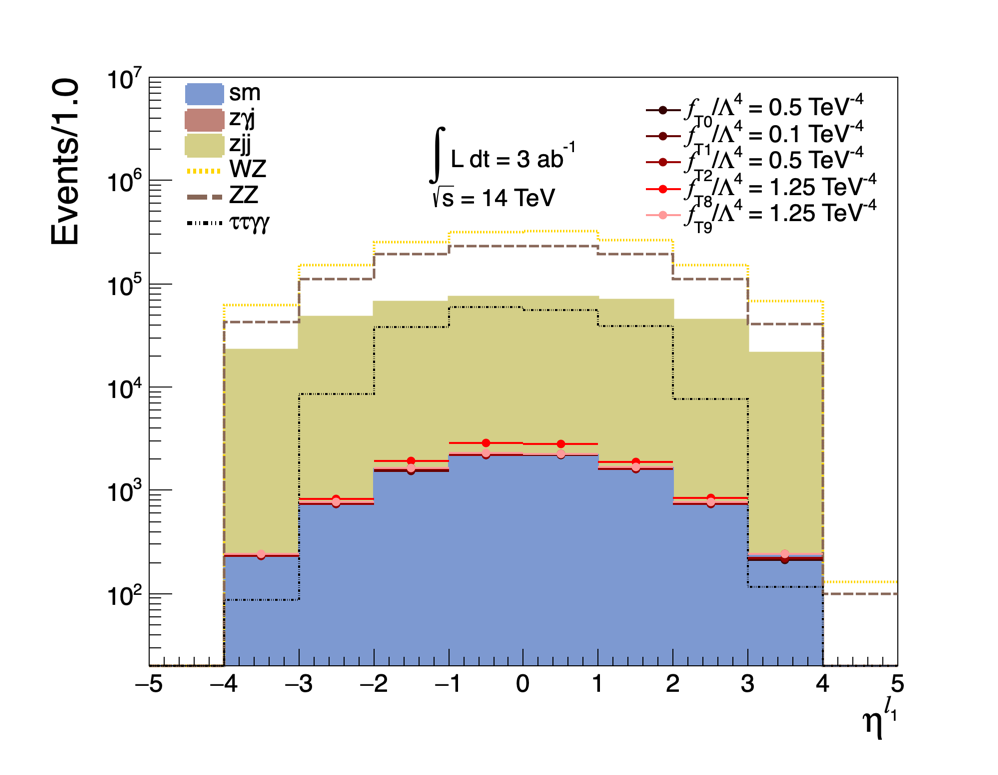}\\\includegraphics[scale=0.23]{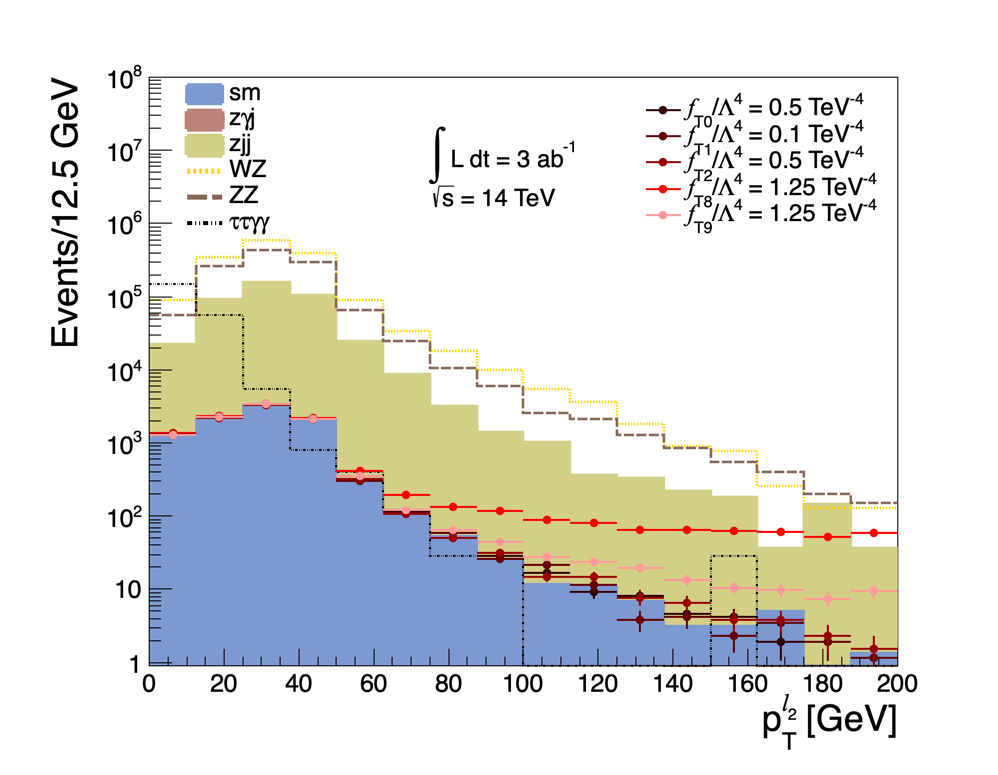}\includegraphics[scale=0.23]{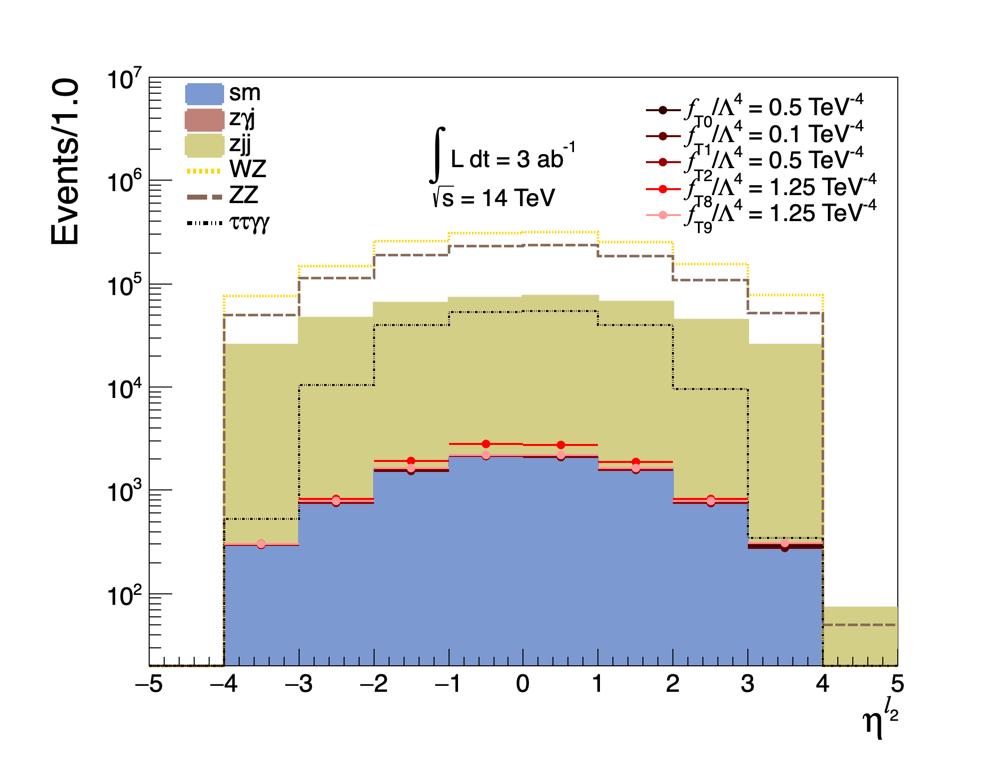}
\caption{Normalized distributions of transverse momentum and pseudo-rapidity of the leading ($l_1$) and sub-leading leptons ($l_2$) after the event selection (Cut-0) for the signals and all relevant backgrounds processes at HL-LHC with $L_{int}$=3 ab$^{-1}$.  \label{fig4}}
\end{figure}
\begin{figure}
\includegraphics[scale=0.23]{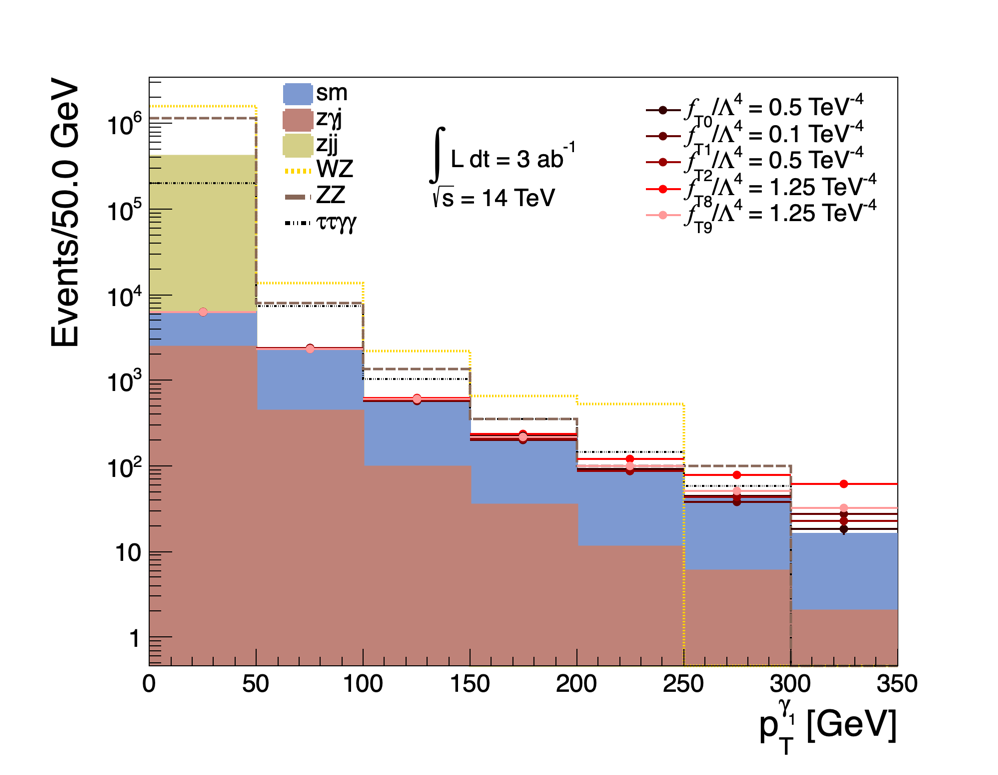}\includegraphics[scale=0.23]{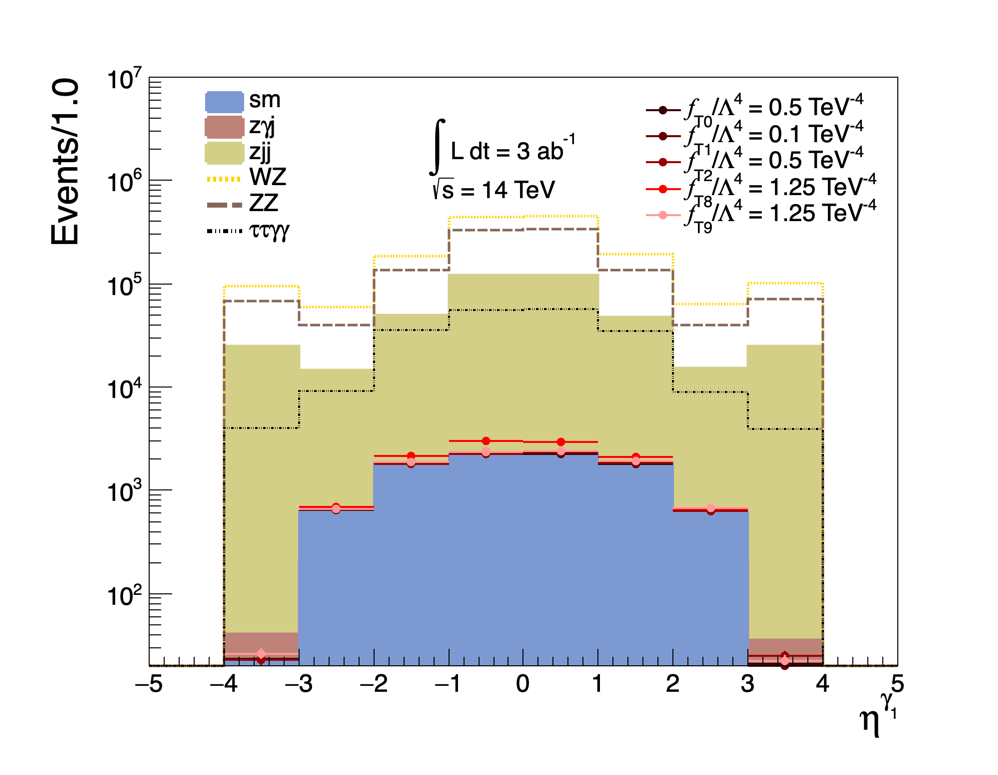}\\\includegraphics[scale=0.23]{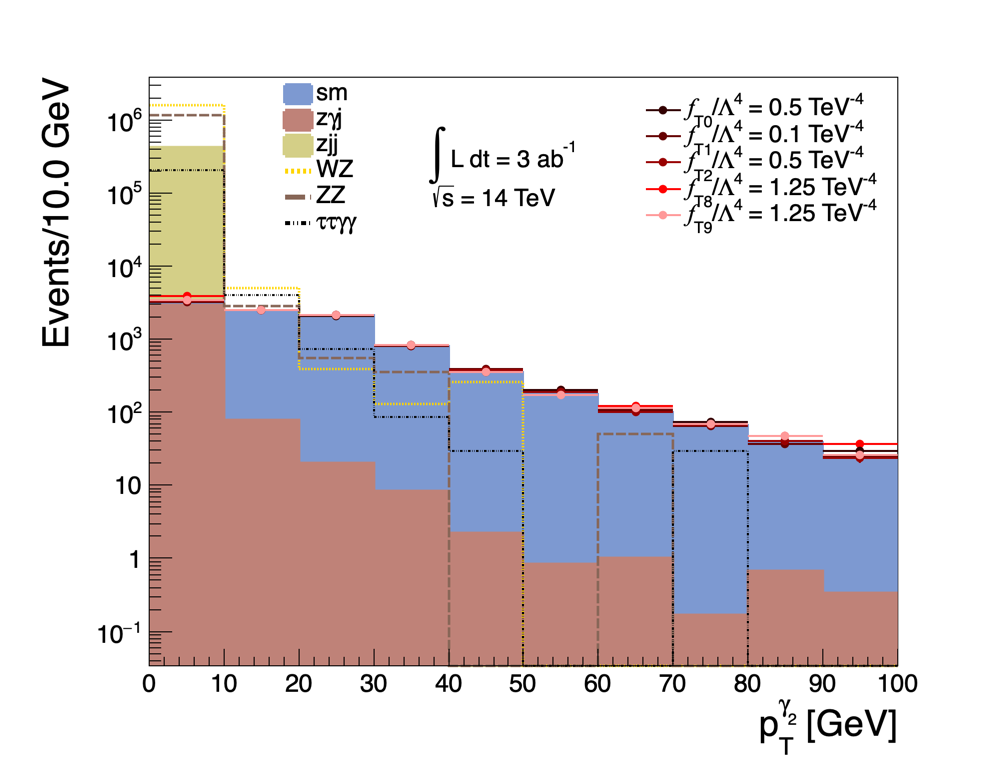}\includegraphics[scale=0.23]{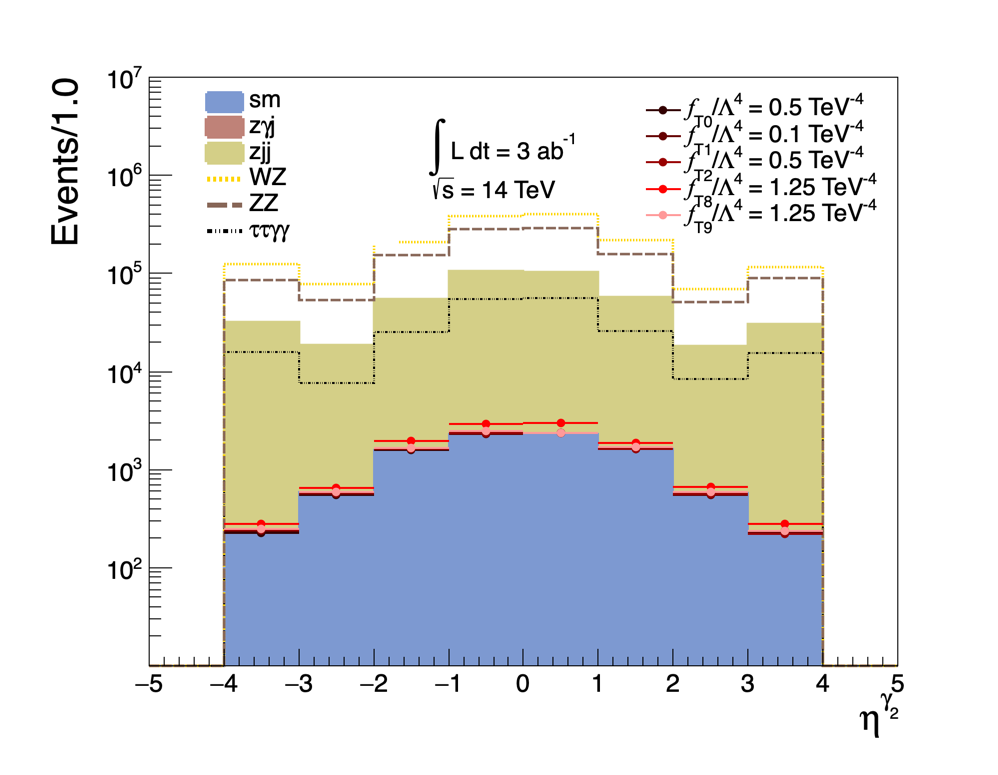}
\caption{Normalized distributions of transverse momentum and pseudo-rapidity of the leading ($\gamma_1$) and sub-leading leptons ($\gamma_2$) after the event selection (Cut-0) for the signals and all relevant backgrounds processes at HL-LHC with $L_{int}=$ 3 ab$^{-1}$. \label{fig5}}
\end{figure}
The normalized distributions of the transverse momentum as well as pseudo-rapidity of leading and sub-leading charged leptons and photons in the selected events (Cut-0) for signal and relevant backgrounds are shown in Fig. \ref{fig4} (Fig.\ref{fig7}) and Fig.\ref{fig5} (Fig.\ref{fig8}) for HL-LHC (FCC-hh) options, respectively.  Similar distribution have also been obtained for HE-LHC. 
By considering typical requirements of the experimental detectors and normalized distributions of final state leptons and photons, we impose minimal cuts on the transverse momentum and rapidity of the final-state charged leptons and photons as given Cut-1 and Cut-2 in Table \ref{tab2} for each collider options. Since the signal event contains two photons, we can safely suppress the event contamination and avoid infrared divergences by using a minimum transverse momentum and pseudo-rapidity cuts of the leading and sub-leading photons for the other background processes. 
\begin{figure}
\includegraphics[scale=0.23]{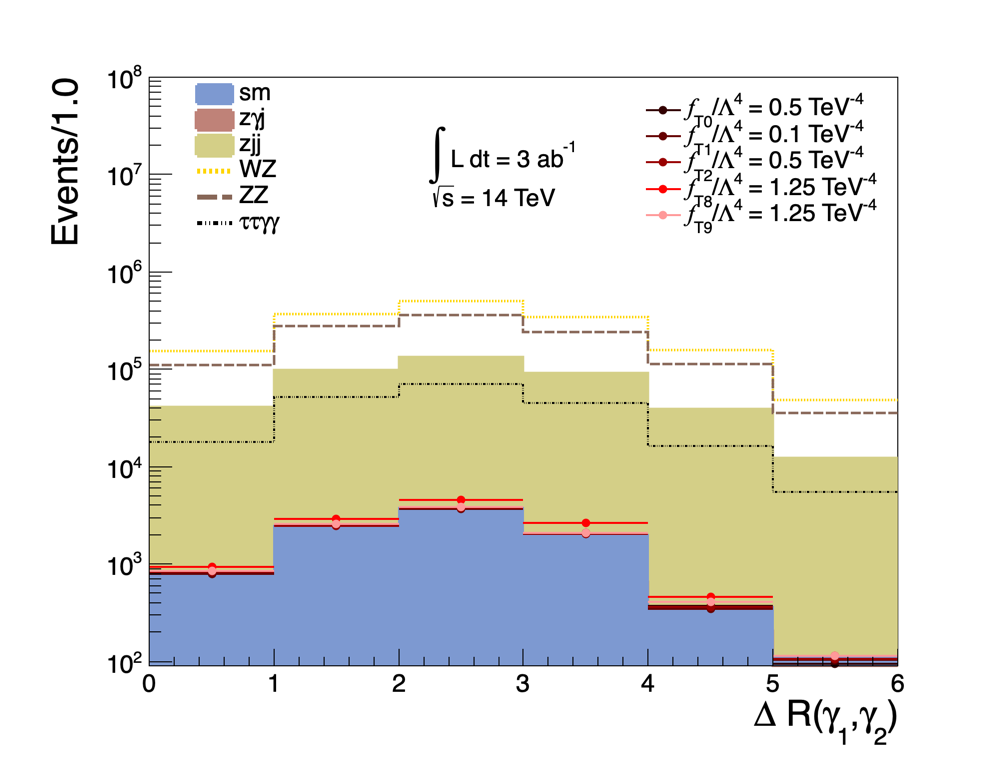}\includegraphics[scale=0.23]{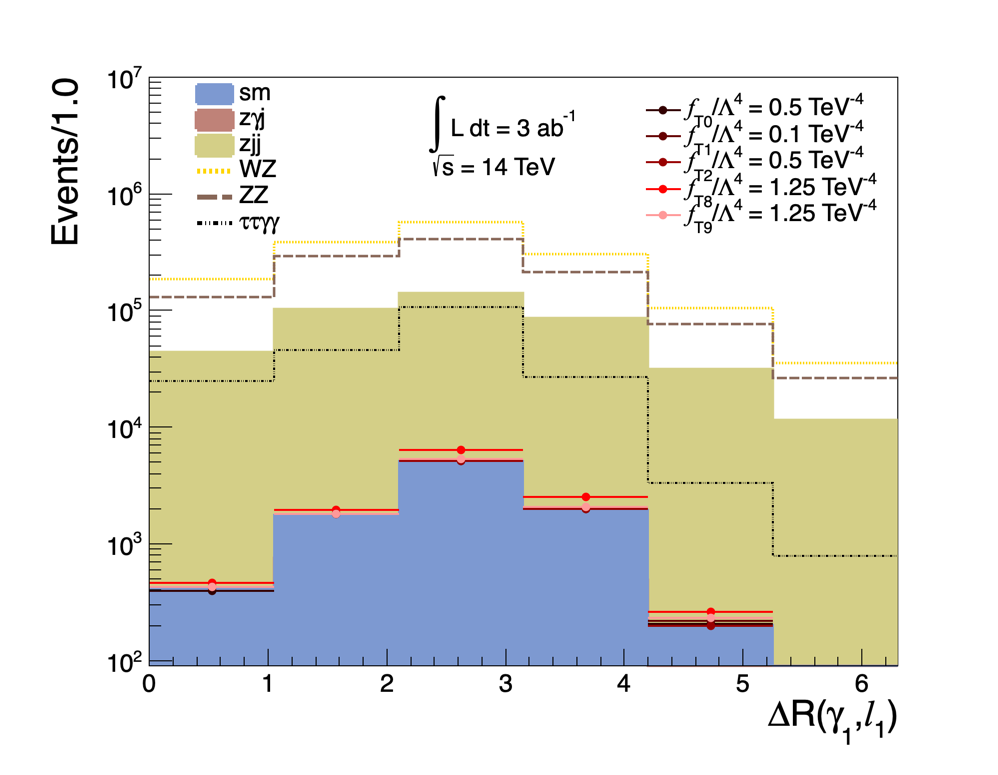}\\\includegraphics[scale=0.23]{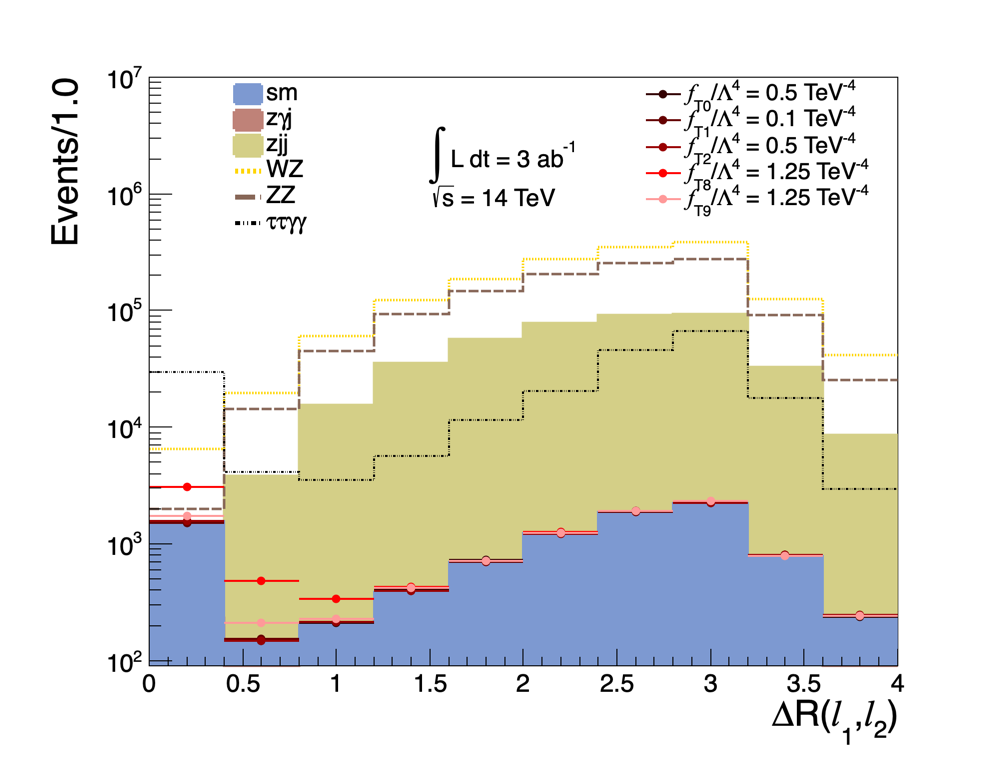}\includegraphics[scale=0.23]{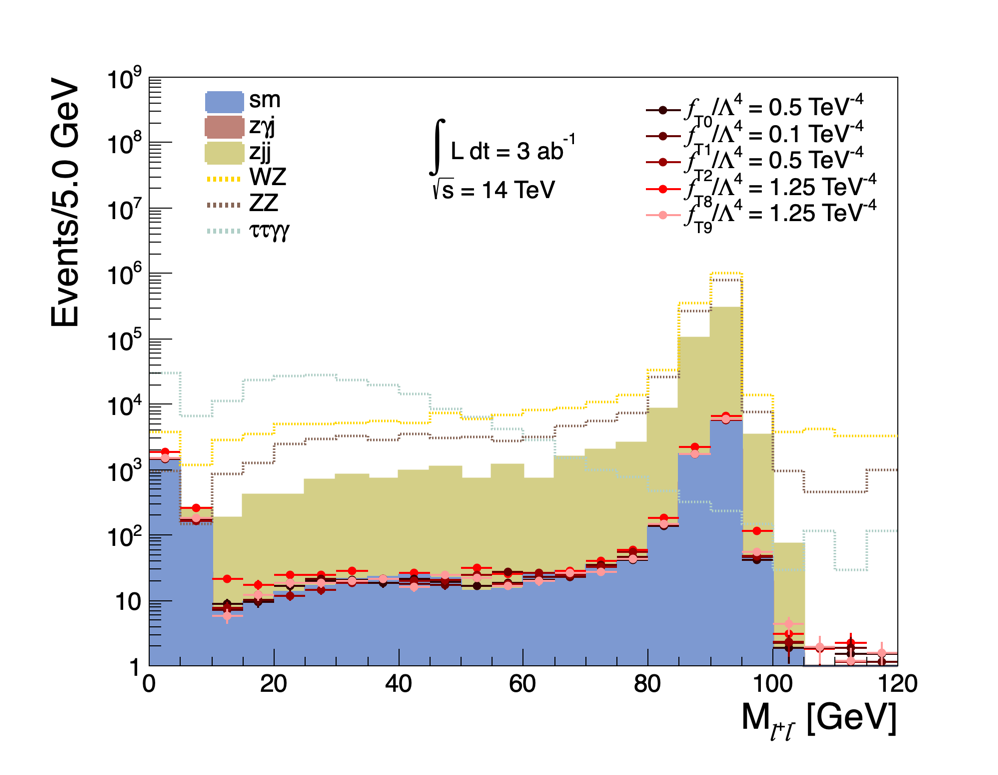}
\caption{Normalized $\Delta R(\gamma_1,\gamma_2)$, $\Delta R(\gamma_1, l_1)$, $\Delta R(l_1, l_2)$  and charged lepton pair invariant mass distribution for 
the signals and all relevant backgrounds processes at HL-LHC with $L_{int} = 3 $ab$^{-1}$ \label{fig6}}
\end{figure}
\begin{figure}
\includegraphics[scale=0.23]{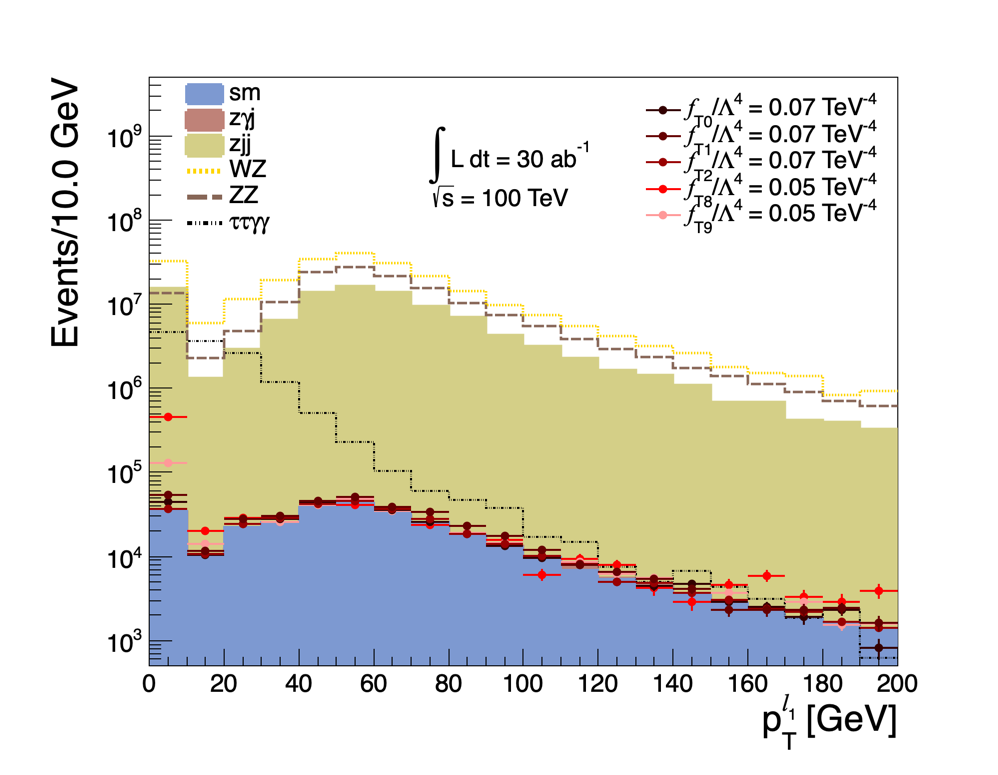}\includegraphics[scale=0.23]{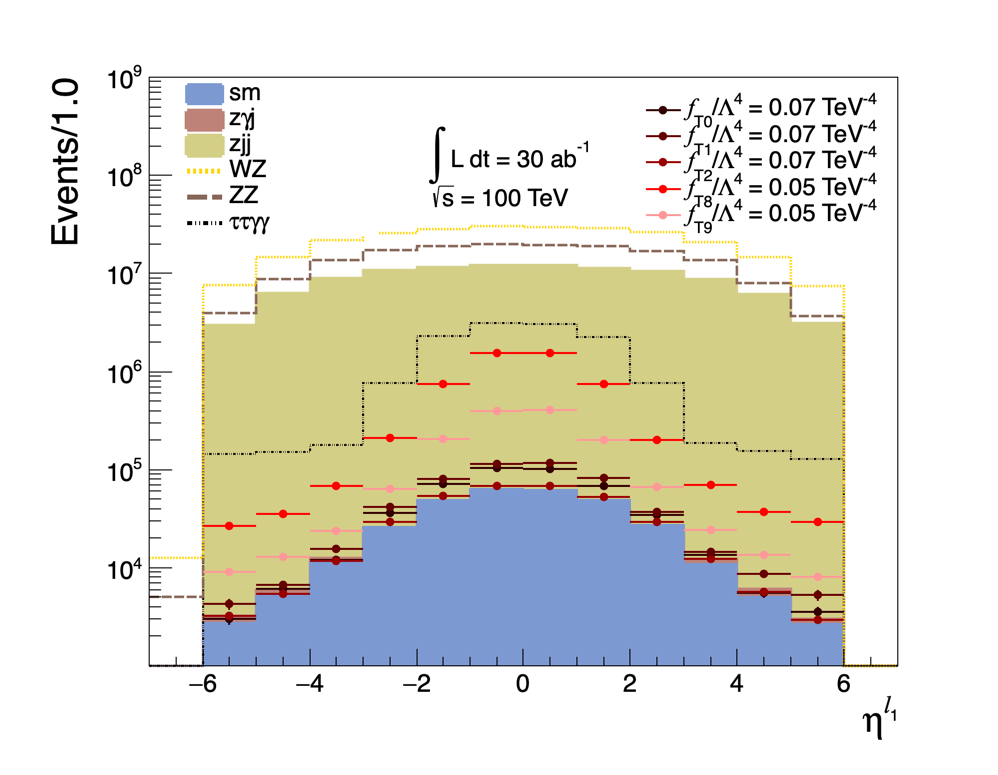}\\\includegraphics[scale=0.23]{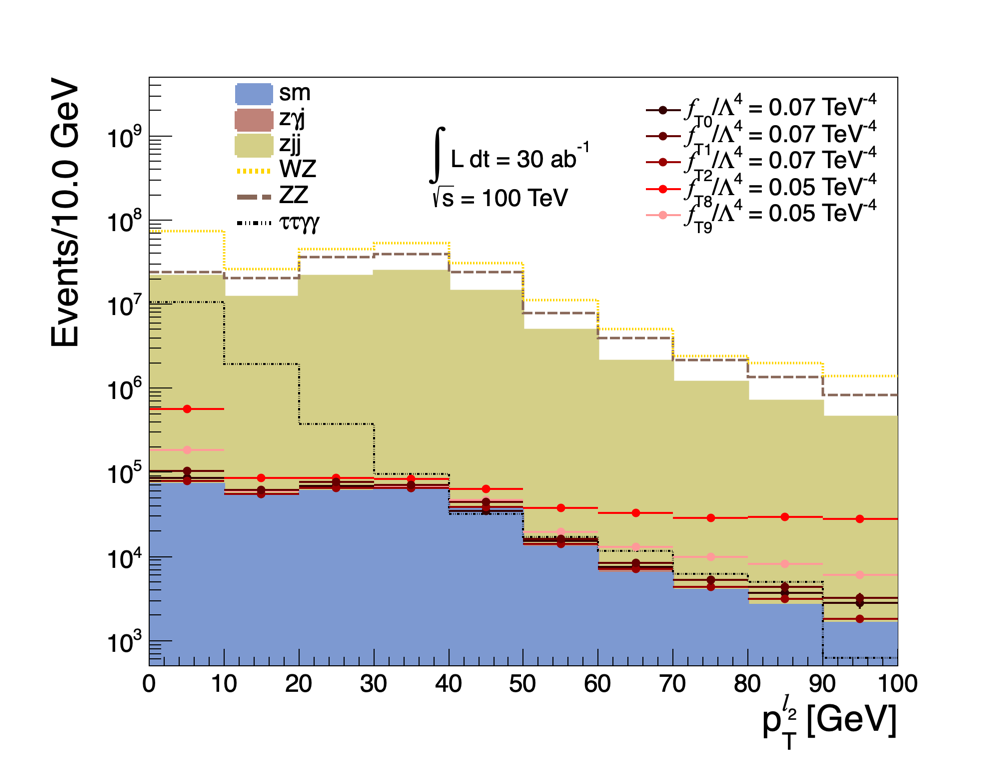}\includegraphics[scale=0.23]{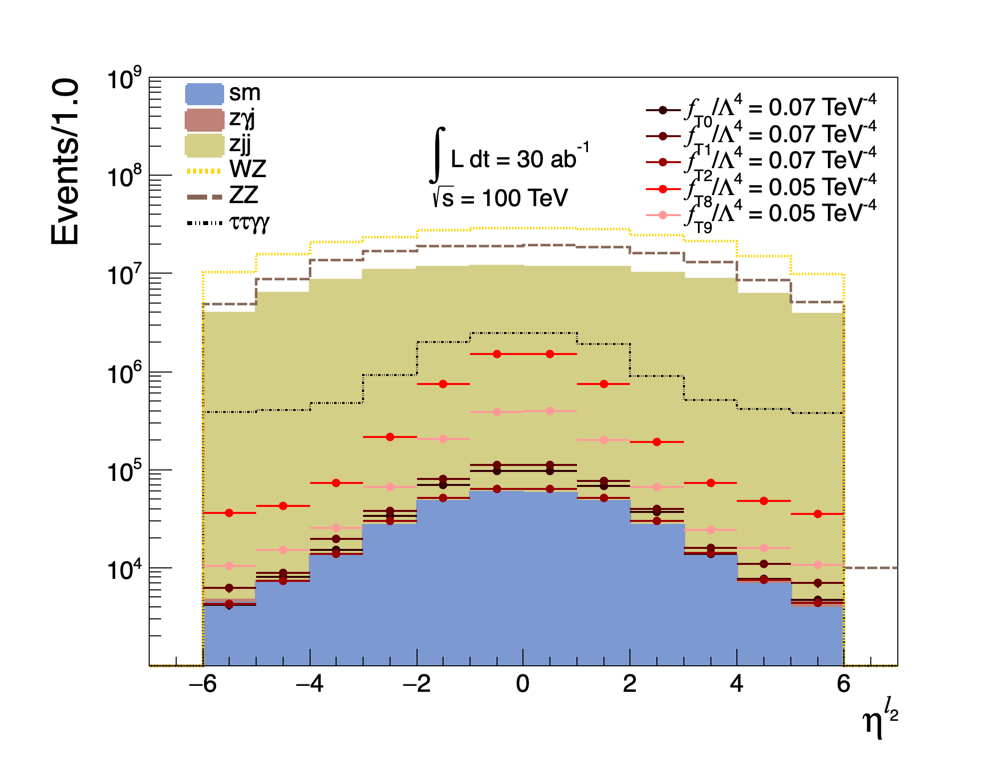}
\caption{Normalized distributions of transverse momentum and pseudo-rapidity of the leading ($l_1$) and sub-leading leptons ($l_2$) after the event selection (Cut-0) for 
the signals and all relevant backgrounds processes at FCC-hh with $L_{int}=30$ ab$^{-1}$. \label{fig7}}
\end{figure}
Furthermore, the normalized distributions of leading and sub-leading photons ($\Delta R(\gamma_1,\gamma_2)$), leading photon and leading charged lepton ($\Delta R(\gamma_1,l_1)$), leading and sub-leading charged leptons ( $\Delta R(l_1,l_2)$) separation in the pseudorapidity-azimuthal angle plane as well as the invariant mass of the oppositely signed charged lepton pair are given in Fig.\ref{fig6} (Fig.\ref{fig9}) for HL-LHC (FCC-hh). To have well-separated photons and charged leptons in the phase space that leads to be identified separate objects in the detector, we require separations as $\Delta R(\gamma_1,\gamma_2)$> 0.4, $\Delta R(\gamma_1,l_1)$> 0.4 and $\Delta R(l_1,l_2)$< 1.4 (Cut-3). We also impose  the invariant mass window cut around the Z boson mass peak as 81 GeV $ < M_{l^+l^-} < $ 101 GeV (Cut-4) to suppresses the virtual photon contribution to the di-lepton system. Since requiring the high transverse momentum photon eliminates the fake backgrounds, we plot the transverse momentum of leading photon for HL-LHC, HE-LHC and FCC-hh in Fig.\ref{fig10} (left to right ) to define a region which is sensitive to aQGC. From these normalized plots we apply a cut on $p_{T}^{\gamma_1}$ as 160 GeV, 250 GeV and 300 GeV for each collider option, respectively (Cut-5).  The flow of cuts are summarized in Table \ref{tab2} for each hadron-hadron colliders that we analyzed. The normalized number of events after applied cuts are presented in Table \ref{tab3} to demonstrate the effect of the cuts used in the analysis for signal  ($f_{T8}/\Lambda^{4}$=0.7, 0.07 and 0.01 TeV $^{-4}$ chosen for illustrative purpose) and relevant backgrounds. The number of events in this table are normalized to the cross section of each process times the corresponding integrated luminosity for each collider options, $L_{int}$ = 3, 15 and 30 ab$^{-1}$ for the HL-LHC, HE-LHC and FCC-hh, respectively. We apply a fake-rate of jet-photon $\epsilon_{j \rightarrow \gamma}= 0.0007\cdot e^{-p_{T}[GeV]/187}$ for  $Z\gamma j$ and $Zjj$ backgrounds as discussed previously. As it can be noticed in this table, all other relevant backgrounds are reduced more than signal and $sm$ background. Overall cut efficiencies of our study are 3.6\% and 1.3\%  for signal and $sm$ background at HL-LHC option while they are 1.1\% and  0.64\% at HE-LHC and 1.5\% and 0.63\% at FCC-hh option.
\begin{table}
\caption{The cumulative number of events for signal ( $f_{T8}/\Lambda^{4}$=0.7, 0.07 and 0.01 TeV $^{-4}$ ) and relevant background processes after applied cuts. The numbers are normalized to the cross section of each process times the integrated luminosity, $L_{int}$ = 3 ab$^{-1}$, 15 and 30 ab$^{-1}$ for the HL-LHC, HE-LHC and FCC-hh respectively. The number of events between the parenthesis are obtained after applying UV bounds for signals. We apply a fake-rate of jet-photon $\epsilon_{j \rightarrow \gamma}= 0.0007\cdot e^{-p_{T}[GeV]/187}$ for  $Z\gamma j$ and $Zjj$ backgrounds. \label{tab3}}
\begin{ruledtabular}
\begin{tabular}{cccccccc}
Cuts &signal&sm&$Z\gamma j$&$Zjj$&$WZ$&$ZZ$&$\tau \tau \gamma \gamma$ \\ \hline
&$f_{T8}/\Lambda^{4}$=0.7 TeV $^{-4}$&&&HL-LHC  \\ \hline
Cut-0&10012  (9869)&9502&3027&$420661$&1599830&1161880&209514 \\
Cut-1&5277  (5117) &4914&1610&$212264$&871430&640538&6860\\
 Cut-2 &3528 (3385)&3293&29&78892&392&600&87\\
Cut-3 &420 (378) &168&2&0&0&0&0\\
Cut-4 & 381 (353)&128&1&0&0&0&0\\
Cut-5& 366  (230)&124&1&0&0&0&0\\\hline
&$f_{T8}/\Lambda^{4}$= 0.07 TeV $^{-4}$&&&HE-LHC  \\ \hline
Cut-0&84234 (84144)&83075&35191&$5768510$&$17932500$&$13334500$&1828320 \\
Cut-1&43192 (43113)&42920&18435&$2785550$&90355710&6905490&65364\\
 Cut-2 &18817 (18744)&18724&162&510&0&5666&0\\
Cut-3 &2554 (2483)&2369&36&0&0&1700&0\\
Cut-4 & 2448 (2376)&2290&28&0&0&567&0\\
Cut-5& 682 (609)&516&4&0&0&0&0\\\hline
&$f_{T8}/\Lambda^{4}$=0.01 TeV $^{-4}$&&&FCC-hh  \\ \hline
Cut-0&339659 (336093)&316474&277688&$104987000$&$255968000$&$162914000$&$13170400$\\
Cut-1&135863 (132902)&125512&108020&$29593500$&$66414700$&$52153500$&280985\\
 Cut-2 &84834 (82976)&79066&793&0&24973&14963&0\\
Cut-3 &19272 (17494)&16211&198&0&12486&0&0\\
Cut-4 &18467 (16848)&15585&149&0&12486&0&0\\
Cut-5&5205 (4010)&2958&20&0&0&0&0\\
\end{tabular}
\end{ruledtabular}
\end{table}
\begin{figure}
\includegraphics[scale=0.23]{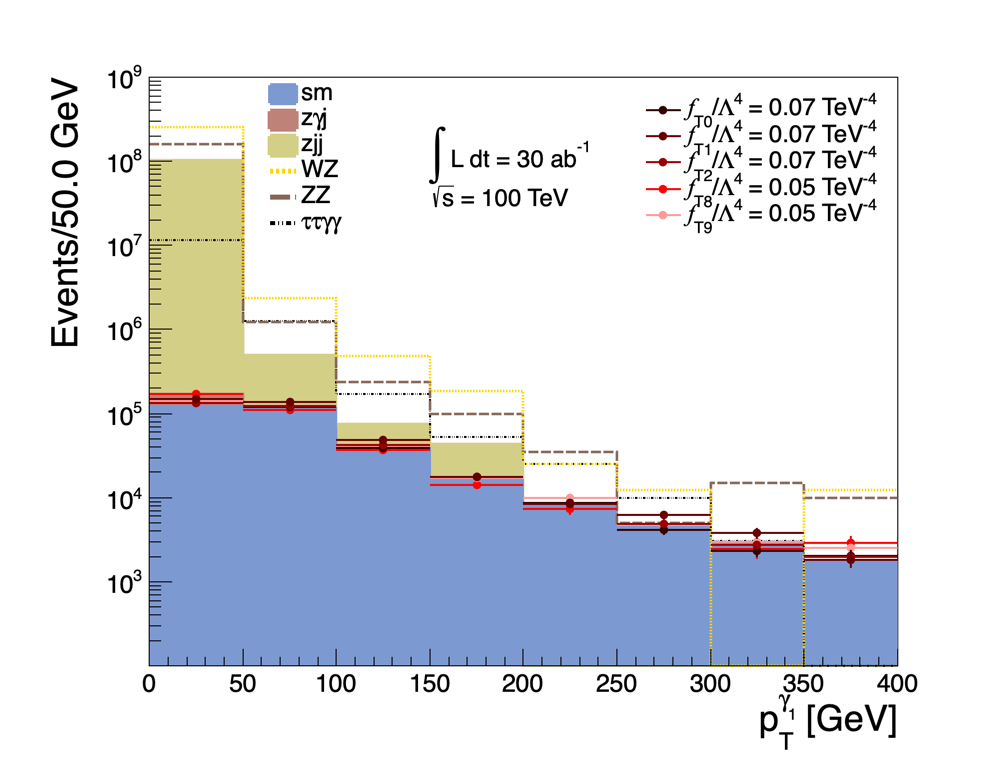}\includegraphics[scale=0.23]{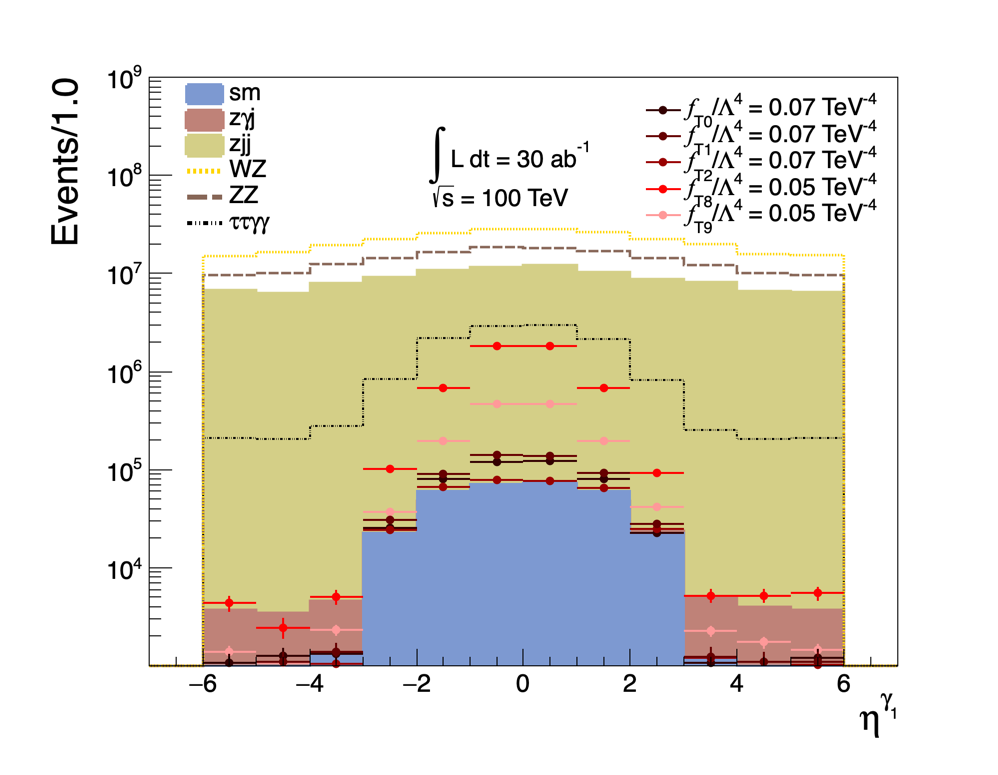}\\\includegraphics[scale=0.23]{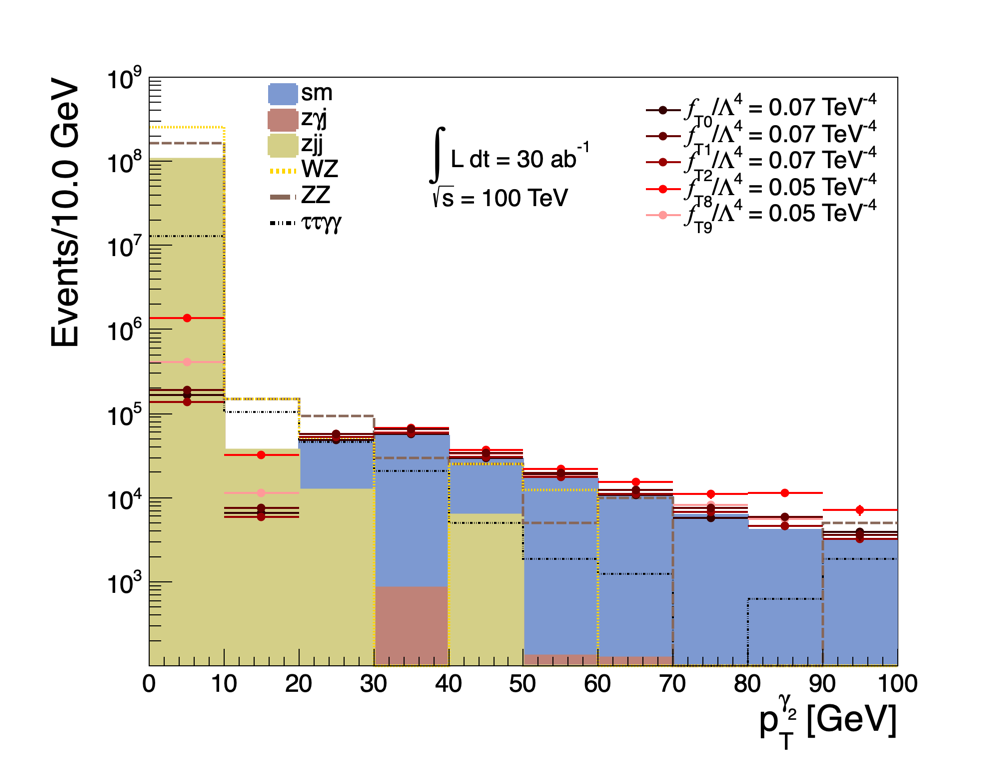}\includegraphics[scale=0.23]{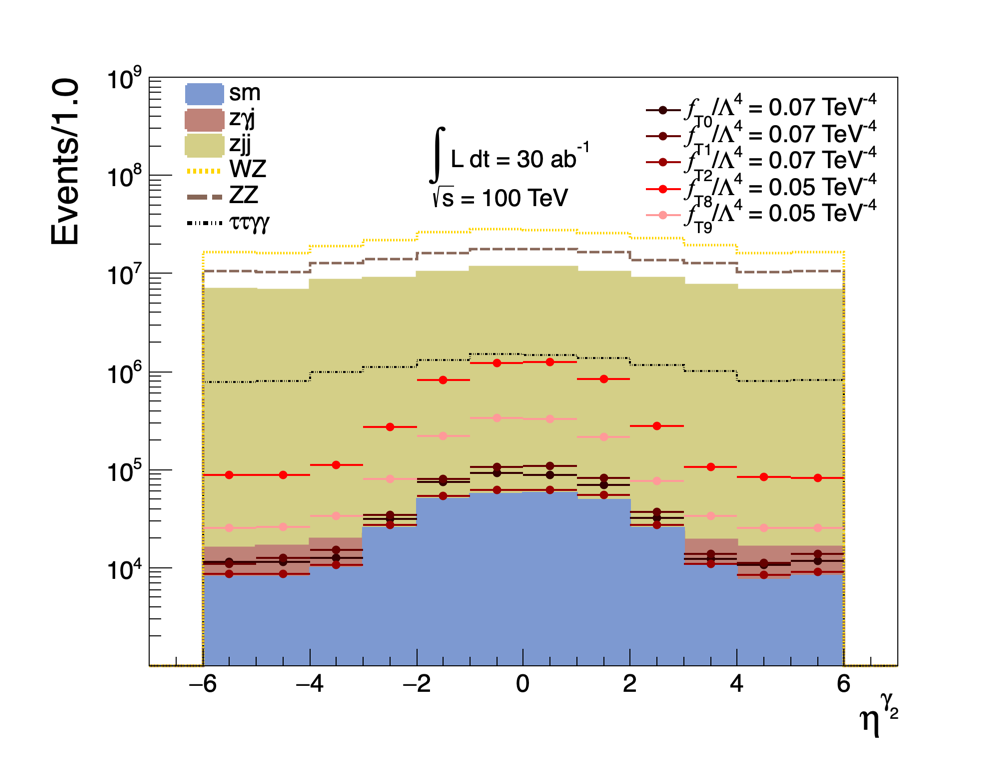}  
\caption{Normalized distributions of transverse momentum and pseudo-rapidity of the leading ($\gamma_1$) and sub-leading leptons ($\gamma_2$) after the event selection(Cut-0) for the signals and all relevant backgrounds processes at FCC-hh with $L_{int}$ = 30 ab$^{-1}$ \label{fig8}}
\end{figure}
\begin{figure}
\includegraphics[scale=0.23]{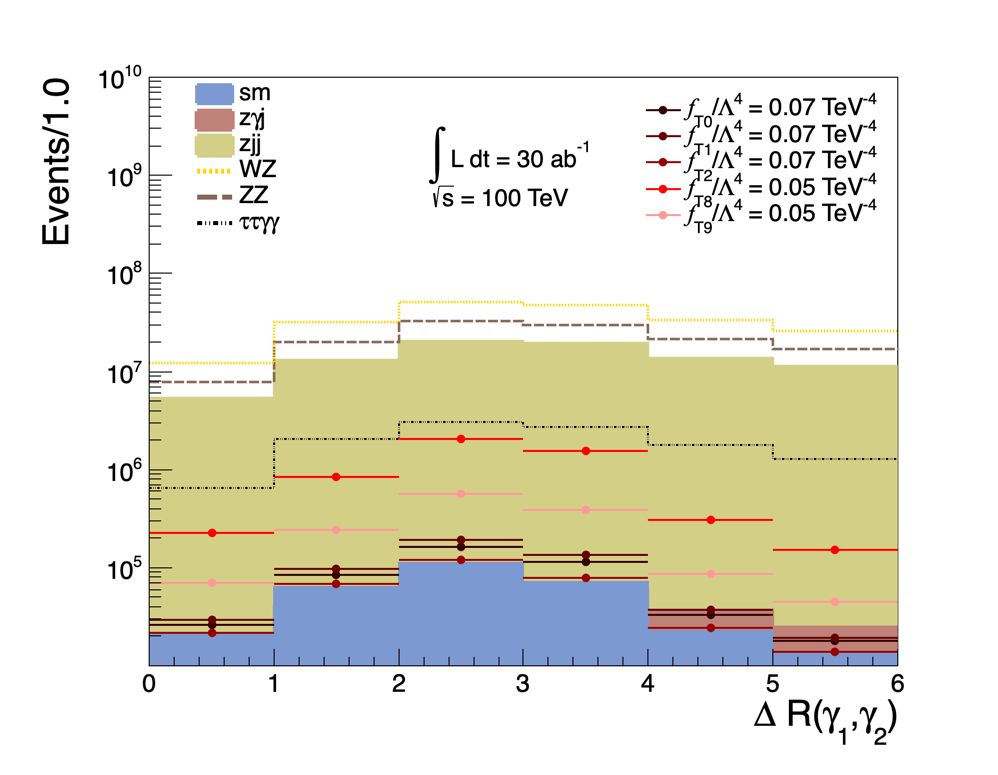}\includegraphics[scale=0.23]{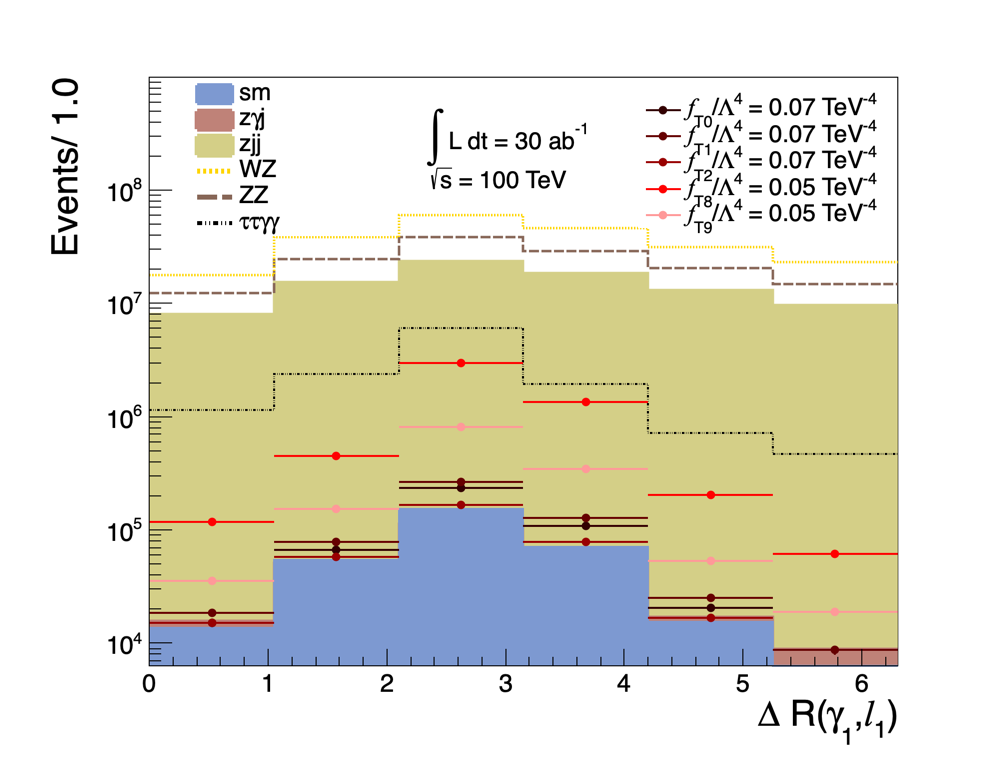}\\\includegraphics[scale=0.23]{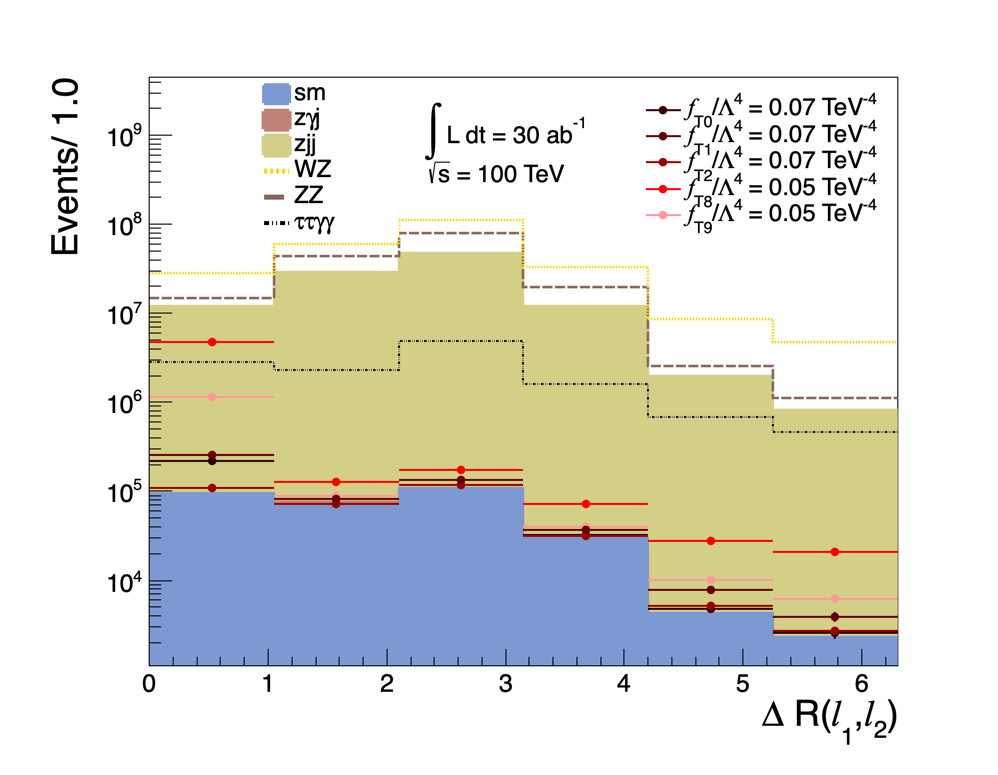}\includegraphics[scale=0.23]{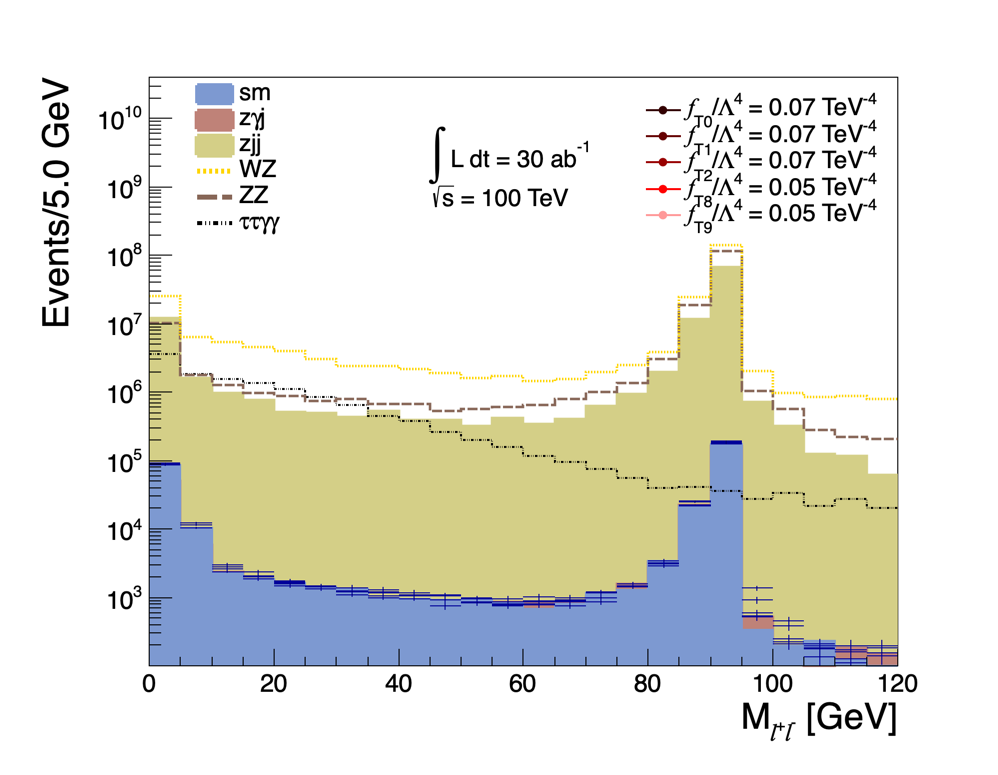}
\caption{Normalized $\Delta R(\gamma_1,\gamma_2)$, $\Delta R(\gamma_1, l_1)$, $\Delta R(l_1, l_2)$ and charged lepton pair invariant mass distribution after the event selection (Cut-0)  for the signals and all relevant backgrounds processes at FCC-hh with $L_{int}$ = 30 ab$^{-1}$. \label{fig9}}
\end{figure}
\begin{figure}
\includegraphics[scale=0.16]{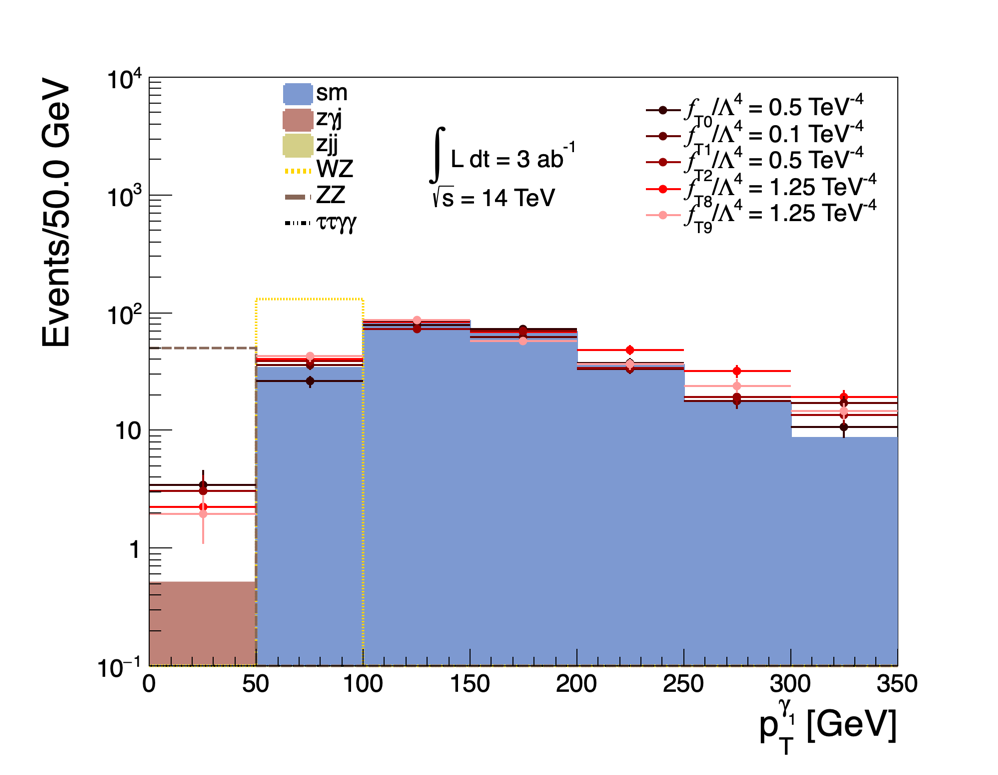}\includegraphics[scale=0.16]{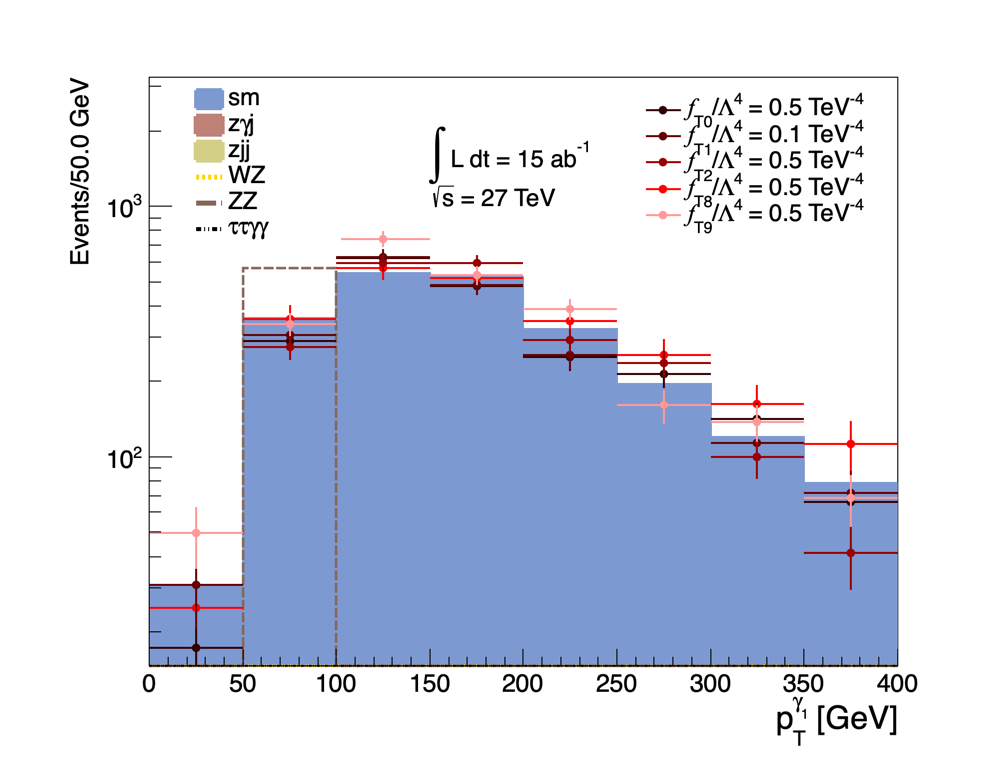}\includegraphics[scale=0.16]{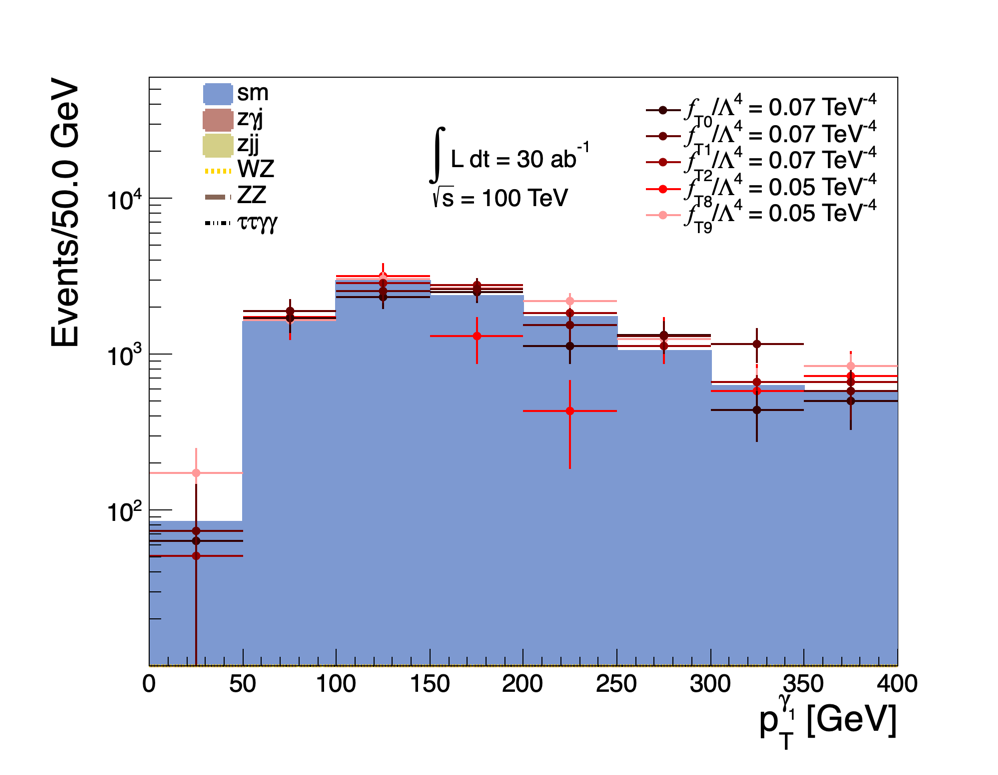}
\caption{Normalized distributions of transverse momentum of the leading photon ($\gamma_1$) after Cut-4 for the signals and all relevant backgrounds processes at HL-LHC, HE-LHC and FCC-hh with $L_{int}$ = 3 ab$^{-1}$, 15 and 30 ab$^{-1}$, respectively \label{fig10}}
\end{figure}
\section{Sensitivity of anomalous quartic gauge couplings for HL-LHC, HE-LHC and FCC-hh }\label{secIV}
\begin{figure}
\includegraphics[scale=0.23]{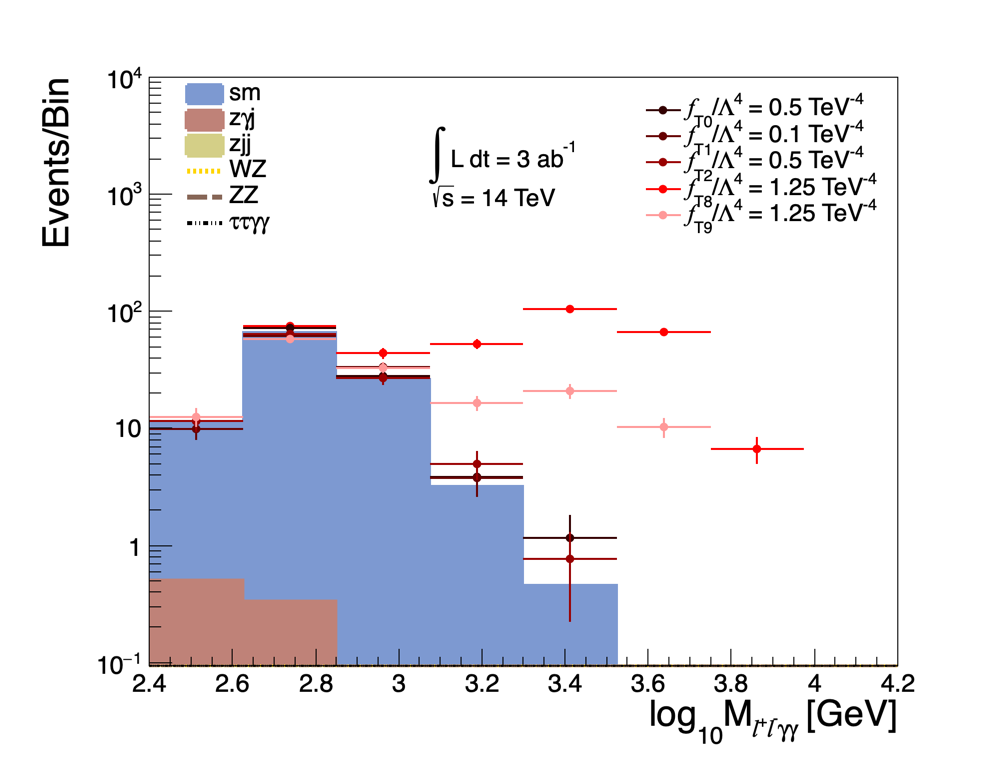}\includegraphics[scale=0.23]{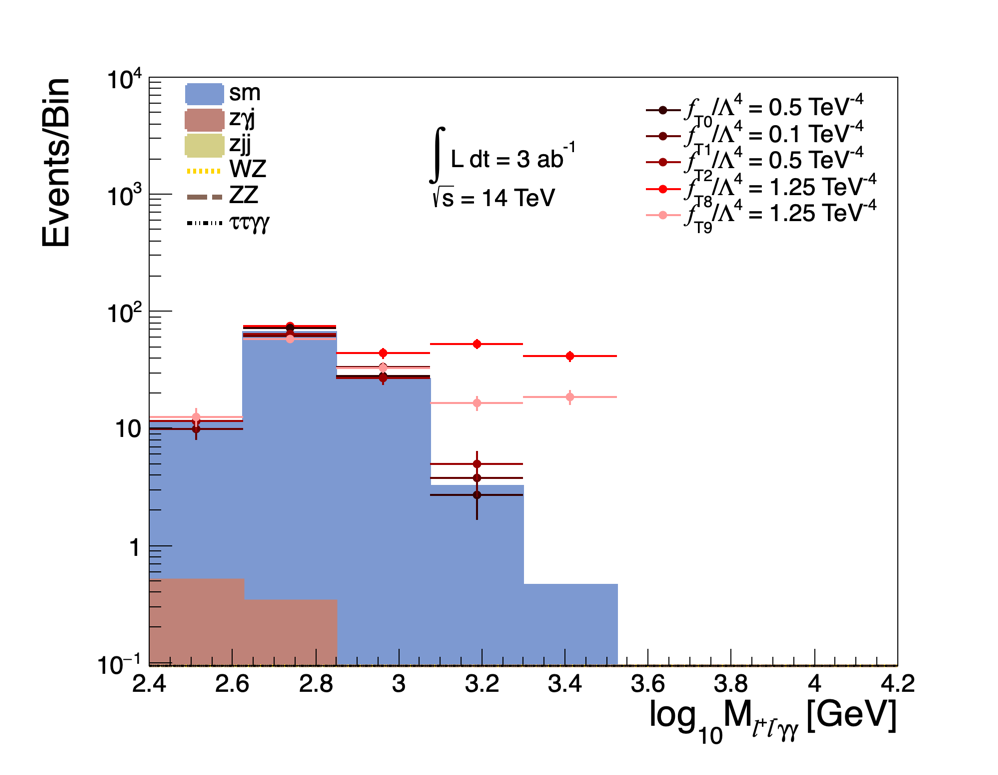}\\\includegraphics[scale=0.23]{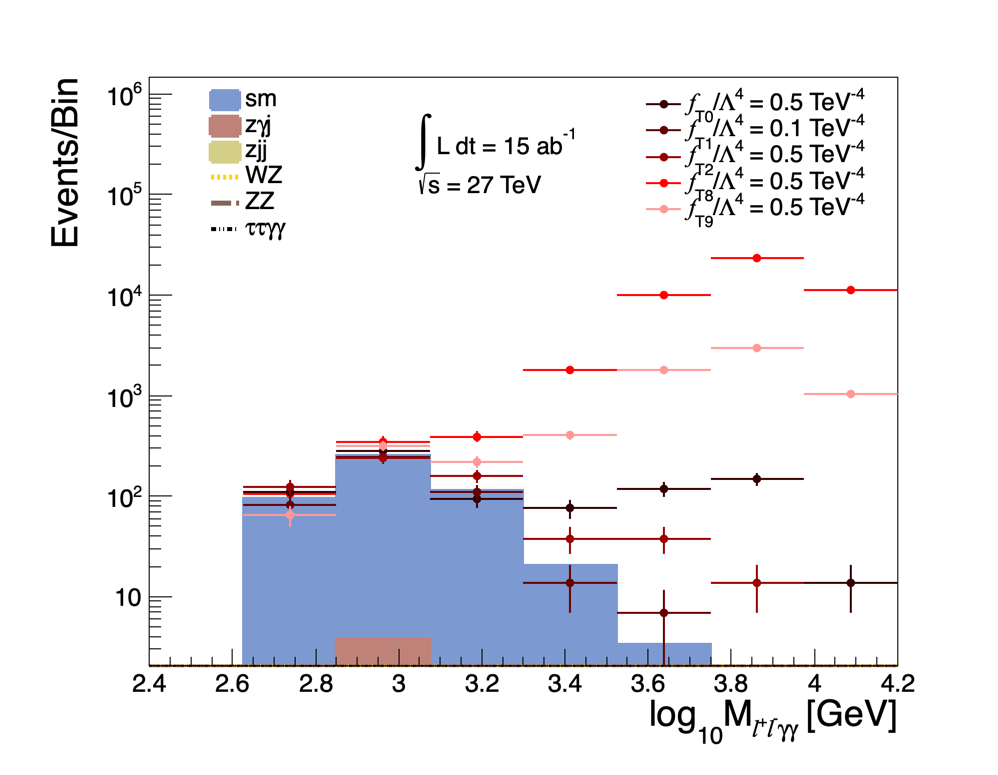}\includegraphics[scale=0.23]{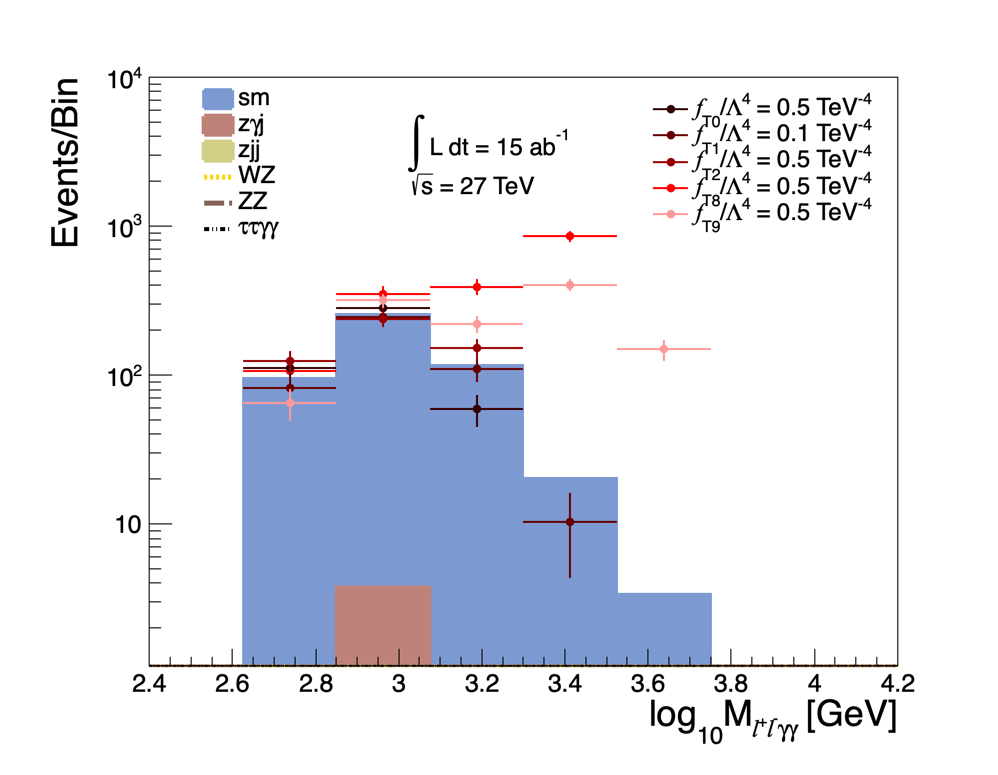}\\\includegraphics[scale=0.23]{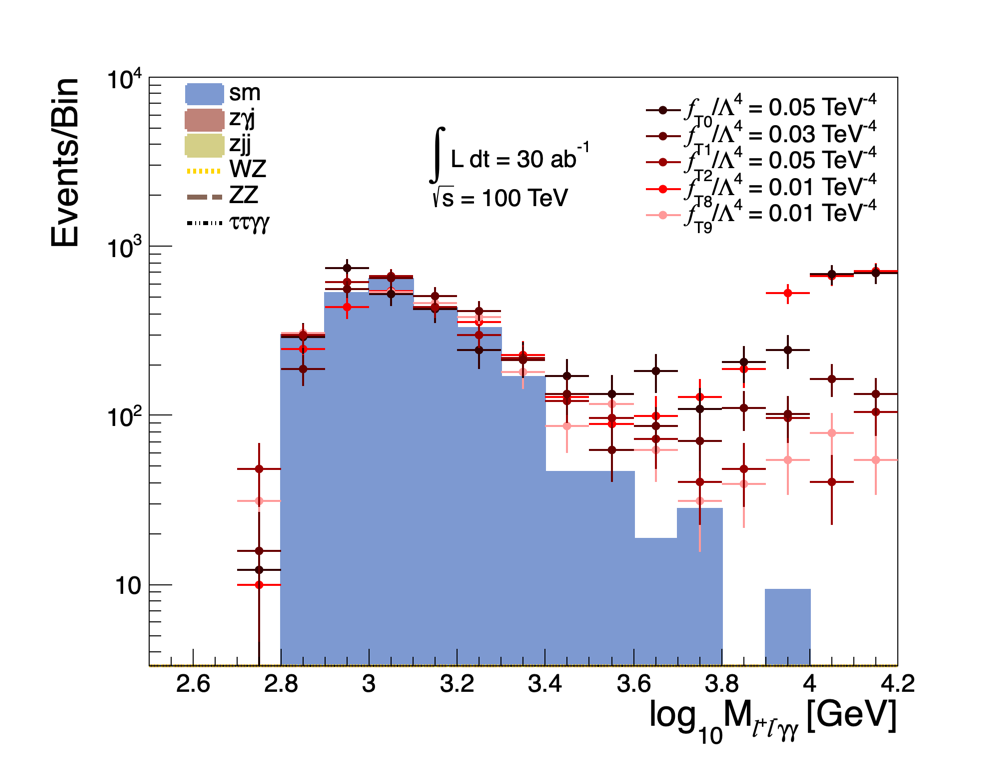}\includegraphics[scale=0.23]{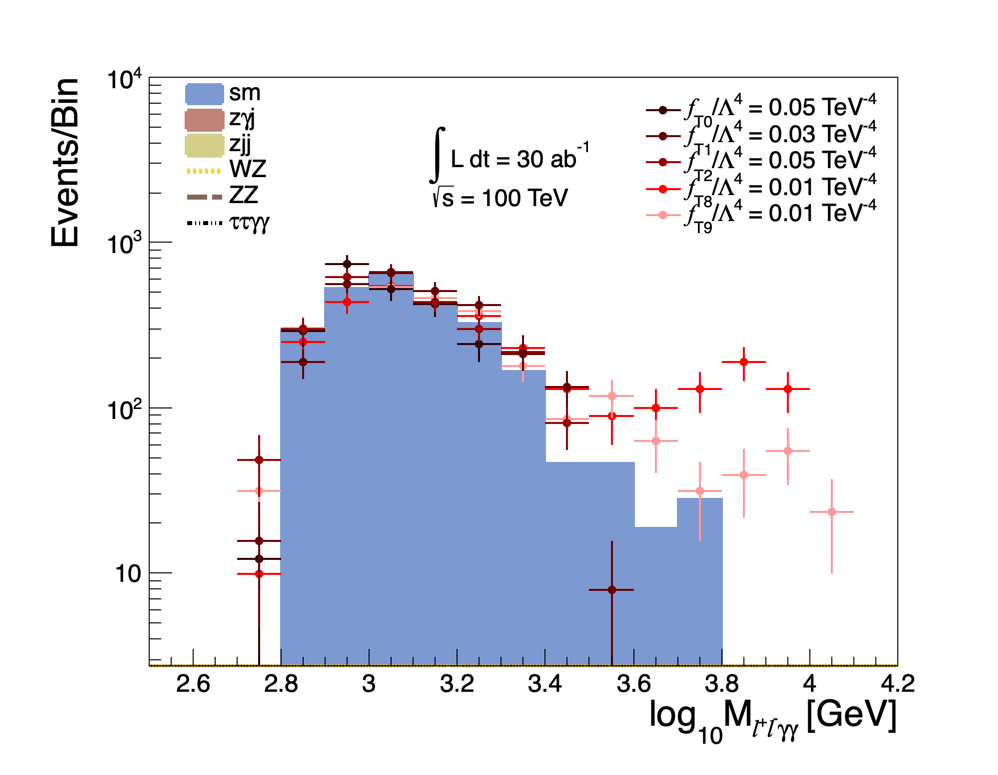}
\caption{Reconstructed mass  of the leading and sub-leading charged leptons and photons after cut-5 without the UV bound (left panles) and with the UV bound (right panels) for the HL-LHC, HE-LHC and FCC-hh  with an integrated luminosity of  $L_{int}$ = 3 ab$^{-1}$, 15 ab$^{-1}$ and 30 ab$^{-1}$, respectively.\label{fig11}}
\end{figure}

In this section, we discuss the detail of the method to obtain sensitivity bounds on anomalous quartic gauge couplings, and then determine the bounds on $f_{T0}/\Lambda^4$, $f_{T1}/\Lambda^4$, $f_{T2}/\Lambda^4$, $f_{T8}/\Lambda^4$ and $f_{T9}/\Lambda^4$ with an integrated luminosity $L_{int}$ = 3 ab$^{-1}$, 15 ab$^{-1}$, 30 ab$^{-1}$ for HL-LHC, HE-LHC and FCC-hh, respectively. The distribution of the reconstructed 4-body invariant mass of $l^+l^-\gamma\gamma$ system for HL-LHC, HE-LHC and FCC-hh options given in Fig. \ref{fig11} (top to bottom, respectively). is used to constrain the aQGC parameters under the hypothesis of absence of anomalous triple gauge couplings. 
The contribution of aQGCs is more visible at high values of the reconstructed 4-body invariant mass of $l^+l^-\gamma\gamma$ system. The effect of UV bound is also presented in this figure where the distributions with (without) UV bound is on the right (left).  The UV bounds are determined by Fig.\ref{fig1} where it is related to the strength of the aQGCs (e.g., the UV bound values for $f_{T8}/\Lambda^{4}$ coupling values of 1.25 TeV$^{-4}$, 0.5 TeV$^{-4}$ and 0.01 TeV$^{-4}$ are 2.5 TeV, 3.2 TeV and 8.4 TeV). As seen from Fig.\ref{fig11}, the applied UV bounds impose an upper-cut in the invariant mass of the $l^+l^-\gamma\gamma$ system which guarantees that the unitarity constraints are always satisfied. The number of events after applying UV bounds for signals are given in the parenthesis at Table \ref{tab3} for comparison purpose.
 

In order to obtain a continuous prediction for the anomalous quartic gauge couplings after Cut-5, a quadratic fit is performed to number of events for each couplings ($\sum_i^{n_{bins}} N_{i}^{NP}$) obtained by integrating the invariant mass distribution of $l^+l^-\gamma\gamma$ system in Fig.\ref{fig11}. The obtaining of the 95\% Confidence Level (C.L.) limit on a one-dimensional aQGC parameter is performed by $\chi^2$  test which corresponds to 3.84

\begin{eqnarray}
\chi^{2} =\sum_i^{n_{bins}}\left(\frac{N_{i}^{NP}-N_{i}^{B}}{N_{i}^{B}\Delta_i}\right)^{2}
\end{eqnarray}

where $N_i^{NP}$ is the total number of events in the existence of aQGC, $N_i^B$ is total number of events of the corresponding SM backgrounds in $i$th bin, $\Delta_i=\sqrt{\delta_{sys}^2+\frac{1}{N_i^B}}$ is the combined systematic ($\delta_{sys}$) and statistical uncertainties in each bin. 

\begin{figure}
\includegraphics[scale=0.44]{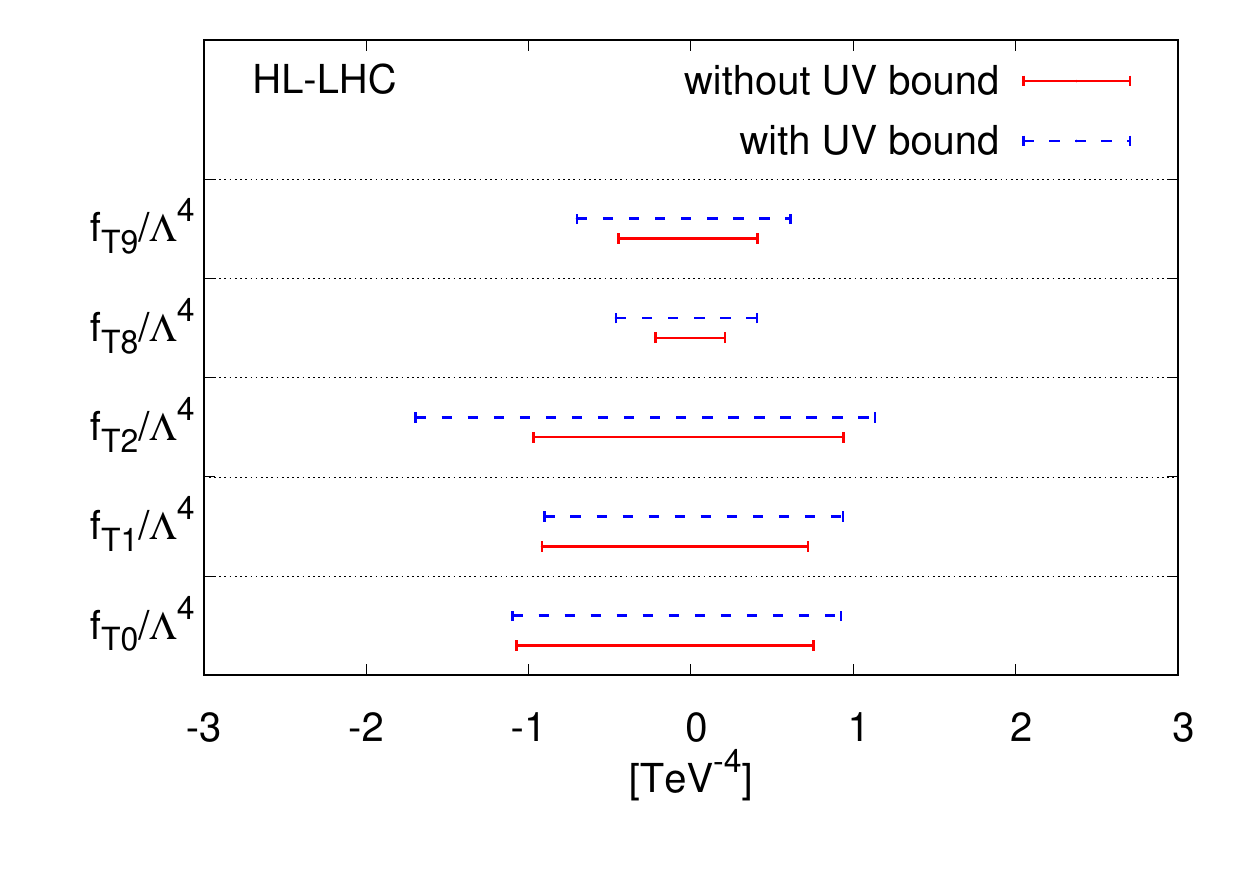}\includegraphics[scale=0.44]{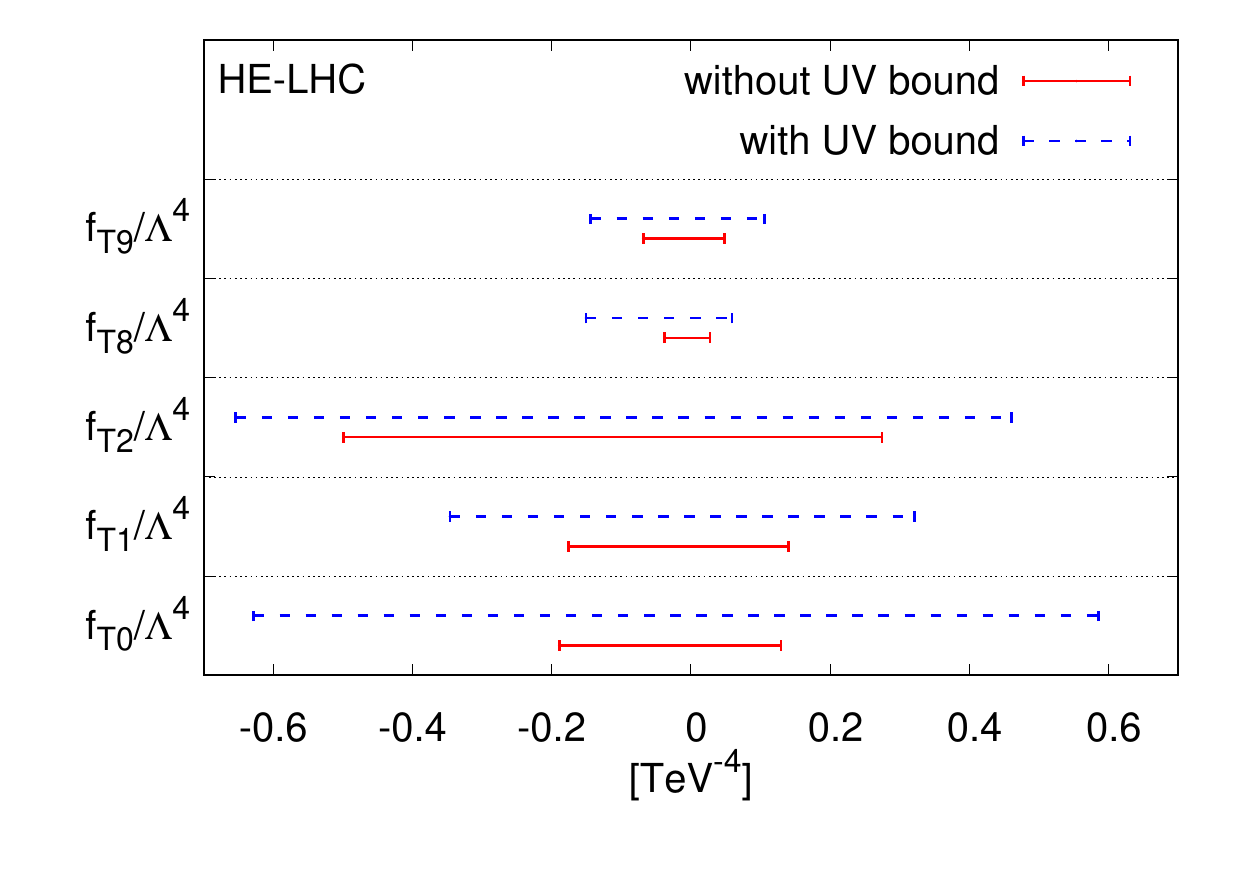}\includegraphics[scale=0.44]{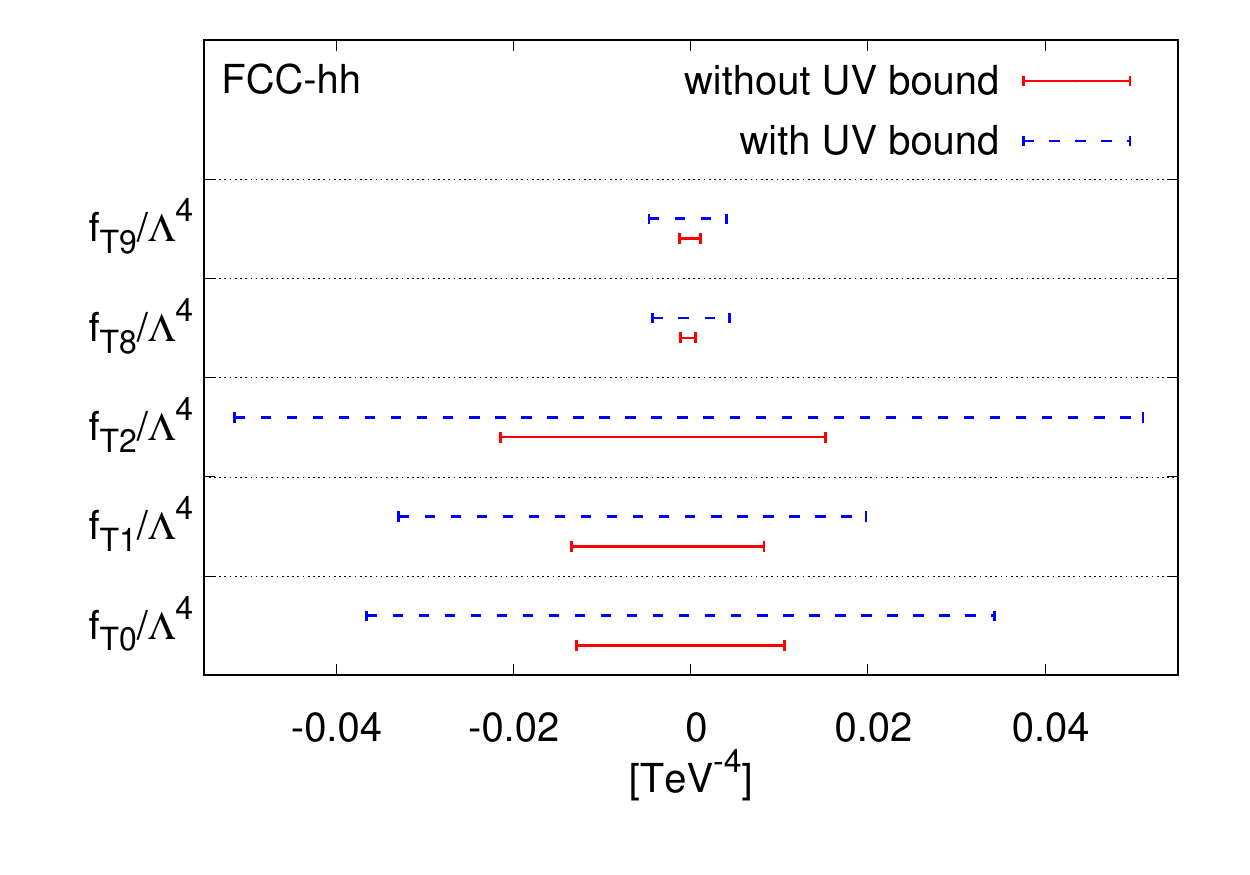}
\caption{95\% C.L. limits on anomalous quartic gauge couplings with and without the unitarity bounds ($\Lambda_{UV}$) considering $\delta_{sys}$=0 of systematic errors with an integrated luminosity of $L_{int}$ = 3 ab$^{-1}$, 15 ab$^{-1}$ and 30 ab$^{-1}$ for the HL-LHC, HE-LHC and FCC-hh, respectively. \label{fig12}}
\end{figure}

The sensitivities of the aQGCs can exhibit in two ways; 
i) allowing the collider to probe sensitivity to new physics above the unitary bound by measuring the direct generation of new resonances and their couplings; ii) assuming that the new physics is probed "virtually" by high-dimension operators that only contain SM fields, and ensuring that the generated events remain below the unitary bounds in the di-boson mass.
The 95\% C.L. limits of the aQGCs obtained by varying the coefficients of one nonzero operator at a time without UV bound are given in Table \ref{tab4} for HL-LHC, HE-LHC and FCC-hh collider options. This table also presents limits with $\delta_{sys}$=  0, 3\%, 5\% and 10\% systematic errors as well as the unitarity bounds defined as the scattering energy at which the aQGC coupling strength is set equal to the observed limit. Following the second way discussed above, we present obtained limits at 95\% C.L. with UV bound applied on the invariant mass distribution of $l^+l^-\gamma\gamma$ system for HL-LHC, HE-LHC and FCC-hh collider options without systematic errors in Fig. \ref{fig12} where the impact of the UV bound  can be seen. The limits on the aQGCs with UV bounds get worsen as expected since the interference of the SM amplitude with the dimension-eight operators suppresses the square contribution of new physics amplitude.

We point out that our numerical results for the case $f_{T8}/\Lambda^{4}$ and $f_{T9}/\Lambda^{4}$  at HL-LHC agree with that of tri-boson production with an upgraded ATLAS detector at a HL-LHC with an integrated luminosity of 3000 fb$^{-1}$ at a center-of-mass energy of 14 TeV \cite{ATLAS:2013uod}. In terms of comparing our results with existent ones, we obtain [-0.22;0.21] and [-0.45;0.41] TeV $^{-4}$ for $f_{T8}/\Lambda^4$ and $f_{T9}/\Lambda^4$  while  the projection limits of ATLAS collaboration are twice as much as our HL-LHC  limits. As shown in Table IV, our results on $f_{T0}/\Lambda^{4}$, $f_{T1}/\Lambda^{4}$ and $f_{T2}/\Lambda^{4}$ ($f_{T8}/\Lambda^{4}$ and $f_{T9}/\Lambda^{4}$ ) for FCC-hh without systematic errors are rather striking as compared to the best experimental results obtained by CMS with 13 TeV data via different production channel \cite{CMS:2019qfk,CMS:2020fqz,CMS:2021gme} by one (two) order of magnitude. In the case of HE-LHC, we obtained one order better limits on $f_{T8}/\Lambda^4$ and $f_{T9}/\Lambda^4$ couplings while similar sensitivities on $f_{T0}/\Lambda^{4}$, $f_{T1}/\Lambda^{4}$ and $f_{T2}/\Lambda^{4}$ couplings compared to the experimental limits obtained from CMS collaboration. Considering the results of the CMS and ATLAS collaboration via the same production channel used in our analysis \cite{ATLAS:2016qjc,CMS:2021jji}, our obtained limits for all aQGC are one to three order of magnitude better as can be seen from the comparison of Table \ref{tab1} and Table \ref{tab4} for HL-LHC, HE-LHC and FCC-hh, respectively. 
The 95\% C.L. limits on $f_{T0}/\Lambda^{4}$ ($f_{T1}/\Lambda^{4}$ ) couplings from $W^+W^-jj$ analysis for HL-LHC, HE-LHC \cite{Azzi:2019yne} and  FCC-hh \cite{Jager:2017owh} are reported as [-0.6,0.6] ([-0.4,0.4]), [-0.027,0.027] ([-0.016,0.016]) TeV $^{-4}$ and [-0.01,0.01] TeV $^{-4}$, respectively. 
Comparing our expected limits for $f_{T0}/\Lambda^{4}$ and $f_{T1}/\Lambda^{4}$ couplings, we obtain limits at the same order for HL-LHC as well as FCC-hh, while one order weaker for HE-LHC. When presenting a realistic physics potential of future colliders for the process $pp\to Z\gamma\gamma$, one should consider the effect of systematic uncertainties on the limits. The source of systematic uncertainties are mainly based on the cross section measurements of $pp\to Z\gamma\gamma$ process with leading-order (LO) or next to leading order (NLO) predictions \cite{Bozzi:2011en} and higher order EW corrections,  the uncertainty in integrated luminosity as well as electrons and jets misidentified as photons. In our study, we focus on LO predictions but do not investigate the impact and validity of these higher-order corrections on the signal and SM background processes. Since the main purpose of this study is not to discuss sources of the systematic uncertainty in detail but to investigate the overall effects of the systematic uncertainty on the limits values of aQGC, we consider three different scenarios of systematic uncertainty. The 95\% C.L. limit values without systematic uncertainties and with three different scenarios of systematic uncertainties as $\delta_{sys}$= 3\%, 5\% and 10\% for three different collider options are quoted in Table \ref{fig4}. The limits on the aQGC considered in this study with systematic error are weaker slightly when realistic systematic error is considered, e.g., compared with a 10\% systematic error and without systematic error, sensitivity of $f_{T9}/\Lambda^4$ gets worsen by about 1.2\%, 1.7\% and 1.5\% for HL-LHC, HE-LHC and FCC-hh, respectively. We expect two (one) order of magnitude better limits on $f_{T8}/\Lambda^4$ and $f_{T9}/\Lambda^4$  ($f_{T0}/\Lambda^{4}$, $f_{T1}/\Lambda^{4}$ and $f_{T2}/\Lambda^{4}$) than the current experimental results even with  $\delta_{sys}$= 10\% systematical uncertainty for FCC-hh whereas comparable limits for $f_{T0}/\Lambda^{4}$, $f_{T1}/\Lambda^{4}$ and $f_{T2}/\Lambda^{4}$ for HE-LHC. 
\section{Summary and Conclusions}\label{secV}
Studying on trilinear and quartic vector boson coupling defined by the $SU(2)_L \times U(1)_Y$ gauge symmetry within the framework of the Standard Model (SM) may lead to an additional confirmation of the model as well as give clues to the existence of new physics at a higher energy scale parameterized with higher-order operators in an Effective Field Theory (EFT). In addition, the possibility of future hadron-hadron colliders with high energy, high luminosity and a high detection expectation access can open a wide window for new physics research for any possible signs of new physics at a higher scale parameterized with higher order operators. In light of these motivations, we focused on the sensitivity of anomalous quartic gauge couplings at HL-LHC, HE-LHC and FCC-hh collider options with corresponding integrated luminosities of $L_{int}$ = 3 ab$^{-1}$, 15 ab$^{-1}$, 30 ab $^{-1}$ via phenomenological analysis of $pp\to Z\gamma\gamma$ production process considering charged lepton decay channel of Z boson. Taking into account the detector effects of these three collider options with the corresponding {\sc Delphes}  card, the detailed analysis of two leptons and two photons final states as well as the  relevant backgrounds is performed using cut-based techniques. In order to determine the region of the phase space that the signal dominates the relevant background, the transverse momentum, pseudo-rapidity, angular $\Delta R$ distributions of the leptons and photons in the final state, as well as the invariant distributions of two leptons with opposite charges and same flavors were investigated for optimum kinematic cuts. The distributions of the leading photon transverse momentum and $l^-l^+\gamma\gamma$ invariant mass system were found to be sensitive indicators of anomalous quartic gauge couplings. Sensitivity of anomalous quartic gauge couplings which one can hope to achieve in future HL-LHC, HE-LHC and FCC-hh collider experiments occurring at $ZZ\gamma\gamma$ and $Z\gamma\gamma\gamma$ vertices for 95\% C.L. has performed with three systematic uncertainty scenario $\delta_{sys}$= 3\%, 5\% and 10\%  using the $l^+l^-\gamma\gamma$ invariant mass system distributions. Since $\mathcal{O}_{T8}$ and $\mathcal{O}_{T9}$ give rise to aQGC containing only the neutral electroweak gauge bosons among the anomalous quartic operators, we reach the remarkable sensitivity on especially $f_{T8}/\Lambda^{4}$ and $f_{T9}/\Lambda^{4}$ couplings for HE-LHC and FCC-hh options comparing with current experimental results as seen from Table \ref{tab4}. If we have compared the current experimental limits with the results determined from this work, sensitivity on the couplings $f_{T8}/\Lambda^{4}$ and $f_{T9}/\Lambda^{4}$ ($f_{T0}/\Lambda^{4}$, $f_{T1}/\Lambda^{4}$ and $f_{T2}/\Lambda^{4}$) for FCC-hh without systematic errors with $L_{int}=$ 30 ab$^{-1}$ are two (one) order better than the current experimental limits obtained from different vector boson scattering process  by CMS collaboration \cite{CMS:2019qfk,CMS:2020fqz,CMS:2021gme}, while the HE-LHC results are one order better limits on $f_{T8}/\Lambda^{4}$ and $f_{T9}/\Lambda^{4}$ couplings but similar sensitivities on $f_{T0}/\Lambda^{4}$, $f_{T1}/\Lambda^{4}$ and $f_{T2}/\Lambda^{4}$ couplings with $L_{int} =$ 15 ab$^{-1}$. Using UV bound on the invariant mass distribution of $l^+l^-\gamma\gamma$ system for HL-LHC, HE-LHC and FCC-hh collider options without systematic errors, obtained limits on aQGCs are strongly affected compared to those without UV bound at 95\% C.L. Further improvements on the limits of aQGCs $f_{Tx}$ can be achieved from analysis of vector boson scattering process at high energy colliders as well as using multivariate techniques.

\begin{table}
\caption{95\% C.L. limits on anomalous quartic gauge couplings in units of TeV$^{-4}$ and the unitarity bounds ($\Lambda_{UV}$) in units TeV  considering $\delta_{sys}$=0, 3$\%$, 5$\%$ and 10\% of systematic errors with an integrated luminosity of $L_{int}$ = 3 ab$^{-1}$, 15 ab$^{-1}$ and 30 ab$^{-1}$ for the HL-LHC, HE-LHC and FCC-hh, respectively.
\label{tab4}}
\begin{ruledtabular}
\begin{tabular}{cc|cc|cc|cc}
&&HL-LHC&&HE-LHC&&FCC-hh&\\ 
&$\delta_{sys}$&Limits [TeV$^{-4}$] & $\Lambda_{UV}$[TeV] &Limits [TeV$^{-4}$]&$\Lambda_{UV}$[TeV]&Limits [TeV$^{-4}$]&$\Lambda_{UV}$[TeV] \\ \hline\hline
                                   &0$\%$&[-1.07;0.76]&     &[-1.89;1.29]$\times10^{-1}$&    &[-1.29;1.05]$\times10^{-2}$&\\
$f_{T0}/\Lambda^{4}$&3$\%$&[-1.10;0.78]&1.4&[-2.04;1.44]$\times10^{-1}$&2.0&[-1.73;1.50]$\times10^{-2}$&3.6 \\
                                   &5$\%$&[-1.14;0.82]&    &[-2.23;1.64]$\times10^{-1}$&    &[-2.11;1.87]$\times10^{-2}$& \\
                                   &10$\%$&[-1.28;0.96]&   &[-2.75;2.16]$\times10^{-1}$&    &[-2.86;2.63]$\times10^{-2}$ \\ \hline
                                   &0$\%$&[-0.92;0.72]&    &[-1.76;1.40]$\times10^{-1}$&    &[-1.34;0.83]$\times10^{-2}$&\\
$f_{T1}/\Lambda^{4}$&3$\%$&[-0.94;0.74]&1.6&[-1.91;1.56]$\times10^{-1}$&2.3&[-1.75;1.23]$\times10^{-2}$&4.5 \\
                                   &5$\%$&[-0.97;0.78]&    &[-2.11;1.76]$\times10^{-1}$&    &[-2.08;1.56]$\times10^{-2}$& \\ 
                                   &10$\%$&[-1.10;0.91]&   &[-2.64;2.29]$\times10^{-1}$&    &[-2.76;2.24]$\times10^{-2}$& \\ \hline
                                   &0$\%$&[-0.97;0.94]&    &[-4.99;2.75]$\times10^{-1}$&    &[-2.15;1.52]$\times10^{-2}$&\\
$f_{T2}/\Lambda^{4}$&3$\%$&[-0.99;0.97]&1.6&[-5.35;3.11]$\times10^{-1}$&2.2&[-2.83;2.20]$\times10^{-2}$&4.3\\
                                   &5$\%$ &[-1.04;1.01]&   &[-5.81;3.57]$\times10^{-1}$&    &[-3.40;2.78]$\times10^{-2}$&\\ 
                                   &10$\%$ &[-1.19;1.15]& &[-7.07;4.82]$\times10^{-1}$&    &[-4.57;3.94]$\times10^{-2}$&\\ \hline
                                   &0$\%$&[-2.19;2.11]$\times10^{-1}$&    &[-3.74;2.77]$\times10^{-2}$&    &[-1.16;0.54]$\times10^{-3}$&\\
$f_{T8}/\Lambda^{4}$&3$\%$&[-2.25;2.17]$\times10^{-1}$&3.9&[-4.04;3.07]$\times10^{-2}$&6.5&[-1.33;0.71]$\times10^{-3}$&17.2 \\
                                   &5$\%$&[-2.35;2.26]$\times10^{-1}$&    &[-4.44;3.47]$\times10^{-2}$&    &[-1.50;0.89]$\times10^{-3}$&\\ 
                                   &10$\%$&[-2.68;2.60]$\times10^{-1}$&   &[-5.51;4.54]$\times10^{-2}$&    &[-1.90;0.13]$\times10^{-3}$&\\ \hline
                                   &0$\%$&[-4.46;4.10]$\times10^{-1}$&    &[-6.83;4.83]$\times10^{-2}$&    &[-1.26;1.09]$\times10^{-3}$&\\
$f_{T9}/\Lambda^{4}$&3$\%$&[-4.58;4.22]$\times10^{-1}$&4.1&[-7.39;5.39]$\times10^{-2}$&6.9&[-1.35;1.18]$\times10^{-3}$&17.7 \\
                                   &5$\%$&[-4.76;4.40]$\times10^{-1}$&    &[-8.11;6.11]$\times10^{-2}$&      &[-1.48;1.31]$\times10^{-3}$&\\ 
                                   &10$\%$&[-5.41;5.05]$\times10^{-1}$&  &[-10.05;8.06]$\times10^{-2}$&     &[-1.84;1.66]$\times10^{-3}$&\\ 
\end{tabular}
\end{ruledtabular}
\end{table}
\newpage
\begin{acknowledgments}
This work was supported by the Scientific and Technological Research Council of Turkey (TUBITAK), Grand No: 120F055. 
\end{acknowledgments}
\appendix
\section{Dimension-eight operators for Anomalous Quartic Gauge Couplings}\label{app}
The anomalous quartic gauge couplings come out  from dimension-8 operators \cite{Eboli:2006wa, Degrande:2013rea, Baak:2013fwa}. There are three classes of operators containing either covariant derivatives of Higgs doublet ($D_\mu \Phi$) only,  or two field strength tensors and two $D_\mu \Phi$, or field strength tensors only;

$i$ ) Two independent operators containing either covariant derivatives of Higgs doublet ($D_\mu \Phi$) only:
\begin{eqnarray}
\mathcal{O}_{S0}&=&[(D_{\mu}\Phi)^{\dag}D_{\nu}\Phi]\times[(D^{\mu}\Phi)^{\dag}D^{\nu}\Phi],\\
\mathcal{O}_{S1}&=&[(D_{\mu}\Phi)^{\dag}D_{\mu}\Phi]\times[(D^{\nu}\Phi)^{\dag}D^{\nu}\Phi],
\end{eqnarray}

$ii$ ) Seven operators are related to $D_{\mu}\Phi$ and the field strength are given by 
\begin{eqnarray}\label{eq2}
\mathcal{O}_{M0}&=&Tr[W_{\mu\nu}W^{\mu\nu}]\times[(D_{\beta}\Phi)^{\dagger}D^{\beta}\Phi],\\
\mathcal{O}_{M1}&=&Tr[W_{\mu\nu}W^{\nu\beta}]\times[(D_{\beta}\Phi)^{\dagger}D^{\mu}\Phi],\\
\mathcal{O}_{M2}&=&[B_{\mu\nu}B^{\mu\nu}]\times[(D_{\beta}\Phi)^{\dagger}D^{\beta}\Phi],\\
\mathcal{O}_{M3}&=&[B_{\mu\nu}B^{\nu\beta}]\times[(D_{\beta}\Phi)^{\dagger}D^{\mu}\Phi],\\
\mathcal{O}_{M4}&=&[(D_{\mu}\Phi)^{\dagger}W_{\beta\nu} D^{\mu}\Phi]\times B^{\beta\nu},\\
\mathcal{O}_{M5}&=&[(D_{\mu}\Phi)^{\dagger}W_{\beta\nu} D^{\nu}\Phi]\times B^{\beta\mu},\\
\mathcal{O}_{M6}&=&[(D_{\mu}\Phi)^{\dagger}W_{\beta\nu}W^{\beta\nu} D^{\mu}\Phi],\\
\mathcal{O}_{M7}&=&[(D_{\mu}\Phi)^{\dagger}W_{\beta\nu}W^{\beta\mu} D^{\nu}\Phi].
\end{eqnarray}
where the field strength tensor of the $SU(2)$ ($W_{\mu\nu}$) and $U(1)$ ($B_{\mu\nu}$) are given by
\begin{eqnarray}
W_{\mu\nu}&=&\frac{i}{2} g \tau^i ( \partial_{\mu} W_{\nu}^i-\partial_{\nu}W_{\mu}^i+g \epsilon_{ijk} W_{\mu}^j W_{\nu}^k)\nonumber\\
B_{\mu\nu}&=&\frac{i}{2}g'(\partial_{\mu}B_{\nu}-\partial_{\nu}B_{\mu}).
\end{eqnarray}
Here, $\tau^i (i=1,2,3)$ are the $SU(2)$ generators, $g=e/sin\theta_W$, $g'=g/cos \theta_W$, $e$ is the unit of electric charge and $\theta_W$ is the Weinberg angle.  

$iii$ ) Eight operators containing field strength tensors only are as follows
\begin{eqnarray}
\mathcal{O}_{T0}&=&\textrm{Tr}[W_{\mu\nu}W^{\mu\nu}]\times \textrm{Tr}[W_{\alpha\beta}W^{\alpha\beta}]\\
\mathcal{O}_{T1}&=&\textrm{Tr}[W_{\alpha\nu}W^{\mu\beta}]\times \textrm{Tr}[W_{\mu\beta}W^{\alpha\nu}]\\
\mathcal{O}_{T2}&=&\textrm{Tr}[W_{\alpha\mu}W^{\mu\beta}]\times \textrm{Tr}[W_{\beta\nu}W^{\nu\alpha}]\\
\mathcal{O}_{T5}&=&\textrm{Tr}[W_{\mu\nu}W^{\mu\nu}]\times B_{\alpha\beta}B^{\alpha\beta}\\
\mathcal{O}_{T6}&=&\textrm{Tr}[W_{\alpha\nu}W^{\mu\beta}]\times B_{\mu\beta}B^{\alpha\nu}\\
\mathcal{O}_{T7}&=&\textrm{Tr}[W_{\alpha\mu}W^{\mu\beta}]\times B_{\beta\nu}B^{\nu\alpha}\\
\mathcal{O}_{T8}&=&[B_{\mu\nu}B^{\mu\nu}B_{\alpha\beta}B^{\alpha\beta}]\\
\mathcal{O}_{T9}&=&[B_{\alpha\mu}B^{\mu\beta}B_{\beta\nu}B^{\nu\alpha}]
\label{eq3}
\end{eqnarray}
A complete list of corresponding quartic gauge boson vertices modified by dimension-8 operators is given in Table \ref{tab5}.
\begin{table}
\caption{Quartic gauge boson vertices modified by the related dimension-8 operator are marked with $\checkmark$ \cite{Baak:2013fwa}.}
\label{tab5}
\begin{ruledtabular}
\begin{tabular}{cccccccccc}
&$WWWW$ &$WWZZ$ &$ZZZZ$  &$WW\gamma Z$ &$WW\gamma\gamma$ & $ZZZ\gamma$ & $ZZ\gamma\gamma$ & $Z\gamma\gamma\gamma$ & $\gamma\gamma\gamma\gamma$ \\ \hline
$\mathcal{O}_{S0},\mathcal{O}_{S1}$ & $\checkmark$& $\checkmark$& $\checkmark$& $$& $$& $$& $$& $$& $$\\
$\mathcal{O}_{M0},\mathcal{O}_{M1},\mathcal{O}_{M6}\mathcal{O}_{M7}$ & $\checkmark$& $\checkmark$& $\checkmark$& $\checkmark$& $\checkmark$& $\checkmark$& $\checkmark$& $$& $$\\
$\mathcal{O}_{M2},\mathcal{O}_{M3},\mathcal{O}_{M4},\mathcal{O}_{M5}$ & $$& $\checkmark$& $\checkmark$& $\checkmark$& $\checkmark$& $\checkmark$& $\checkmark$& $$& $$\\
$\mathcal{O}_{T0},\mathcal{O}_{T1},\mathcal{O}_{T2}$ & $\checkmark$& $\checkmark$& $\checkmark$& $\checkmark$& $\checkmark$& $\checkmark$& $\checkmark$& $\checkmark$& $\checkmark$\\
$\mathcal{O}_{T5},\mathcal{O}_{T6},\mathcal{O}_{T7}$ & $$& $\checkmark$& $\checkmark$& $\checkmark$& $\checkmark$& $\checkmark$& $\checkmark$& $\checkmark$& $\checkmark$\\
$\mathcal{O}_{T8},\mathcal{O}_{T9}$ & $$& $$& $\checkmark$& $$& $$& $\checkmark$& $\checkmark$& $\checkmark$& $\checkmark$\\
\end{tabular}
\end{ruledtabular}
\end{table}

\end{document}